\documentclass[a4paper,11pt]{article}
\pdfoutput=1 
\usepackage{float}
\usepackage{jheppub} 

\usepackage[T1]{fontenc} 
\usepackage{mathtools}
\usepackage{amsmath}
\usepackage{hyperref}

\usepackage{color}
\definecolor{cblue}{RGB}{100,5,255}
\definecolor{cred}{RGB}{255,50,40}
\definecolor{cgreen}{RGB}{40,255,40}
\definecolor{cmagenta}{RGB}{139,0,139}

\providecommand{\xlink}[1]
  {\href{http://arxiv.org/abs/#1}{arXiv:#1}}
\def\bea{\begin{eqnarray}}
\def\eea{\end{eqnarray}}

\title{\boldmath Phenomenological implications of the Friedberg-Lee transformation in a neutrino mass model  with $\mu\tau$-flavored CP symmetry}

\author[a]{Roopam Sinha}
\author[a]{Sukannya Bhattacharya}
\author[b]{Rome Samanta}
\affiliation[a]{Saha Institute of Nuclear Physics, HBNI, Kolkata 700064, India}
\affiliation[b]{Physics and Astronomy,University of Southampton, Southampton, SO17 1BJ, U.K.}
\emailAdd{roopam.sinha@saha.ac.in}
\emailAdd{sukannya.bhattacharya@saha.ac.in}
\emailAdd{romesamanta@gmail.com}

\abstract{We propose a neutrino mass model with $\mu\tau$-flavored CP symmetry, where the effective light neutrino Lagrangian enjoys an additional invariance under a Friedberg-Lee (FL) transformation on the left-handed flavor neutrino fields that leads to a highly predictive and testable scenario. While both types of the light neutrino mass ordering, i.e., Normal Ordering (NO) as well as the Inverted Ordering (IO) are allowed, the absolute scale of neutrino masses is fixed by the vanishing determinant of light Majorana neutrino mass matrix $M_\nu$.  We show that for both types of mass ordering, whilst the atmospheric mixing angle $\theta_{23}$ is in general nonmaximal ($\theta_{23}\neq \pi/4$), the Dirac CP phase $\delta$ is exactly maximal ($\delta=\pi/2,3\pi/2$) for IO and nearly maximal for NO owing to $\cos\delta\propto \sin\theta_{13}$. For the NO, very tiny nonvanishing Majorana CP violation might appear through one of the Majorana phases $\beta$; otherwise the model predicts vanishing Majorana CP violation. Thus, despite the fact, that from the measurement of $\theta_{23}$, it is difficult to rule out the model, any large deviation of $\delta$ from its maximality, will surely falsify the scenario. For a comprehensive numerical analysis, beside fitting the neutrino oscillation global fit data, we also present  a study on the $\nu_\mu\rightarrow \nu_e$ oscillation which is expected to show up Dirac  CP violation in different long baseline experiments. Finally, assuming purely astrophysical sources, we calculate the Ultra  High Energy (UHE) neutrino flavor flux ratios at neutrino telescopes, such as IceCube, from which  statements on the octant of $\theta_{23}$ could be made in our model.}

\begin{document}
\maketitle
\flushbottom

\section{Introduction}In spite of the spectacular developments in last couple of decades, the theoretical origin of neutrino masses, flavor mixing and CP violation\cite{King:2015aea} in the leptonic sector remain unresolved. In addition, models with definitive statements about the mass ordering and the absolute scale of three light neutrino masses are yet to be tested. Experiments so far with solar, atmospheric, reactor and accelerator neutrinos have determined the three mixing angles and the two independent mass-squared differences to a reasonably decent accuracy, while the current cosmological upper bound on the sum of the three light neutrino masses is fairly robust:$\sum_i m_i < 0.17$ eV\cite{Aghanim:2016yuo}. The octant of the atmospheric mixing angle $\theta_{23}$ remains unknown though the best-fit values are reported as $47.2^\circ$ for NO and $48.1^\circ$ for IO\cite{Esteban:2016qun,nufit}. Therefore, a precise prediction of $\theta_{23}$ can be used to exclude and discriminate models in the light of forthcoming precision measurements. On the other hand, the current best-fit values of the Dirac CP phase $\delta$,  are close to $234^\circ$ for NO and $278^\circ$ for IO. While the possibility of CP conservation ($\sin\delta=0$) is allowed at slightly above $1\sigma$, one of the CP violating value $\delta=\pi/2$ is disfavored at 99\% CL. Thus, the remaining CP violating value $\delta=3\pi/2$ and deviations around it still remain potentially viable and tantalizing possibilities. Beside all these, it still remains a baffling conundrum for neutrino experts whether the light neutrinos are Dirac or Majorana in nature. Till date, despite relentless searches, no experimental signature of the neutrinoless double $\beta-$decay signal have been observed. However, the rapid development in the long baseline experiments such as T2K\cite{Abe:2017bay}, NO$\nu$A\cite{Adamson:2017qqn} and also $0\nu\beta\beta$ experiments such as KamLandZen\cite{kam}, GERDA\cite{gerda1,gerda2} is expected to shed light on the above issues shortly. Thus, from a theoretical perspective, this is a moment of paramount importance in neutrino mass model building, since many of the existing models that have predictions of $\theta_{23}$, $\delta$ and the neutrino mass ordering are likely to be challenged through precise measurements of these quantities in ongoing and forthcoming experiments.

Discrete flavor  symmetries\cite{Altarelli:2010gt,Ishimori:2010au,King:2017guk,petcov} have always been the center of attraction in neutrino mass model building scenarios due to their highly testable prediction on neutrino mixing parameters. These include the celebrated $\mu\tau$-interchange symmetry\cite{m1,m2,m3,m4,Fukuyama:1997ky,Fukuyama:2017qxb} which was thought to be dead after the discovery of nonvanishing (now confirmed at more than 5.2$\sigma$\cite{th13}) reactor mixing angle $\theta_{13}$. Interestingly, it has now been resurrected in the neutrino mass models by a simple change of usage. To be precise, by using the $\mu\tau$-interchange symmetry as the generator of a non-standard CP symmetry (${\rm CP}^{\mu\tau}$)\cite{cp1,cp2,cp3}:
\begin{equation}
\nu_{Ll}\to iG_{lm}\gamma^0\nu^C_{Lm},\label{u}
\end{equation}
instead of an exact $\mu\tau$-interchange flavor symmetry:
\begin{equation}
\nu_{Ll}\to G_{lm}\nu_{Lm},\label{a1}
\end{equation}
in the effective neutrino Majorana mass term in the low-energy Lagrangian (density)
 \begin{equation}-\mathcal{L}_{\rm mass}^\nu= \frac{1}{2}\overline{\nu_{Ll}^C} (M_\nu)_{lm}\nu_{Lm} +{\rm  h.c.} .\label{lag}
\end{equation}  Here, $\nu_{Ll}^C=C\overline{\nu_{ Ll}}^T$ and the subscripts $l,m$ spanning the lepton flavor indices $e,\mu,\tau$, while the subscript $L$ denotes left-handed flavor neutrino fields. $M_\nu$ is a complex symmetric matrix ($M_\nu^*\neq M_{\nu}=M_\nu^T$) in lepton flavor space. Though ${\rm CP}^{\mu\tau}$ was proposed few years back\cite{cp4,cp3}, currently it has drawn a lot of attention\cite{cp5,cp6,cp7,cp8,cp9,cp10,cp11,cp12,cp13,cp14,cp15,cp16,cp17,cp18,
cp19,cp20,cp21,cp22,King:2017guk,cp24,Rahat:2018sgs} due to its exact prediction: $\theta_{23}=\pi/4$ and $\delta=\pi/2$ or 3$\pi/2$ (Co-bimaximal mixing\cite{Ma:2015fpa}), which is also a recent hint from T2K\cite{Abe:2017bay}. To make ${\rm CP}^{\mu\tau}$ more predictive, a sizeable body of research has been done combining CP symmetry with other flavor symmetries\cite{King:2017guk}, despite the fact that at very high energy, it is nontrivial to have a consistent theory of CP combined with flavor symmetry\cite{cp7,cp8}.

A particular generalization \cite{cp16,Samanta:2018efa} of \eqref{u} is ${\rm CP}^{\mu\tau\theta}$ which is implemented in the neutrino Majorana mass term with the field transformation \begin{equation}
\nu_{Ll}\to iG^\theta_{lm}\gamma^0\nu^C_{Lm}. \label{cpmutau}
\end{equation} In the neutrino flavor space  $G^{\mu\tau\theta}$ has the generic form
 \begin{equation} G^{\mu\tau\theta}=\begin{pmatrix}-1 & 0 & 0\\0 & -\cos\theta & \sin\theta\\ 0 & \sin\theta & \cos\theta\end{pmatrix}\label{a2},\end{equation}
 with `$\theta$' being an arbitrary mixing angle that mixes the $\nu_{L\mu}$ and $\nu_{L\tau}$ flavor fields. The negative signs in (\ref{a2}) are to comply with the PDG convention.  It is worth noticing that $\theta=\pi/2$ reduces the mixing symmetry $G_{lm}^{\mu\tau\theta}$ to the interchange symmetry $G_{lm}^{\mu\tau}$ and any nonzero value of $\theta-\pi/2$ has the potential to account for the deviation from ${\rm CP}^{\mu\tau}$. Eq.\eqref{a2} is a special case of Eq.8 of Ref.\cite{Chen:2015siy} with $\alpha=\pi,\beta=-\pi$ and $\gamma=0$. Though, in general, ${\rm CP}$ symmetries are highly predictive in terms of mixing angles and CP-violating phases, for most of the cases, it lacks information regarding light neutrino masses and mass ordering unless one invokes additional flavor symmetries to reduce the number of parameters\cite{King:2017guk}, e.g, by the means of `texture zeros' in the light neutrino mass matrix\cite{cp12,cp24}.

In this work, to have  testable predictions  in each sector (masses as well as mixing) instead of any additional flavor symmetry, in combination with \eqref{cpmutau}, we consider a Friedberg-Lee (FL) transformation\cite{fl1,fl2,fl3,fl4,fl5,fl6}
 \begin{equation}\nu_{Ll}\to iG_{lm}^{\mu\tau\theta}\gamma^0\nu^C_{Lm}+\eta_l\xi\label{fl}.
\end{equation} This leads to
\begin{equation}
M^\nu\boldsymbol{\eta}=0,~~\text{and}~~(G^{\mu\tau\theta})^TM_\nu G^{\mu\tau\theta}=M^*_\nu,\label{z1}
\end{equation} where $\eta_l$ ($l=e,\mu,\tau$) are three arbitrary complex numbers, $\boldsymbol{\eta}=(\eta_e~\eta_\mu~\eta_\tau)^T$ and $\xi$ is a fermionic Grassmann field \cite{fl1}. Note that, \eqref{fl} is a simple CP generalization of the ordinary (general) FL transformation (also known as twisted FL symmetry\cite{fl7,fl8})
 \begin{equation}
\nu_{Ll}\to G_{lm}^{\mu\tau\theta}\nu_{L m}+\eta_l\xi
\end{equation} leading to \begin{equation}
M^\nu\boldsymbol{\eta}=0,~~\text{and}~~(G^{\mu\tau\theta})^TM_\nu G^{\mu\tau\theta}=M_\nu.\label{c1}
\end{equation}
We would like to stress that in this work we mainly focus on the effective field transformation (\ref{fl}) and its low energy  phenomenological consequences without an explicit top down model realization like in the cases of CP combined with flavor symmetries \cite{cp10,cp11,cp14}. Nevertheless, the generalized $\mu\tau$ and FL could arise from a discrete flavor symmetries such  $D_4$ \cite{Grimus:2004cc} and singlet scalar extension to the  Standard Model \cite{fl4} respectively. Since the residual symmetries in the charged lepton sector and the neutrino sector decide the low energy predictions for the neutrino parameters, from the phenomenological point of view it is a challenging task to identify proper residual symmetries which are predictive while being consistent with the extant neutrino data.  Individually, flavor symmetries, CP symmetries or FL symmetries would not suffice  to lead to  residual symmetries  which are predictive in mass as well as mixing sectors. That is why certain combinations of these symmetries are always attractive at least at the phenomenological level. For example, various models discussed in \cite{King:2017guk} deal with a combined theory of CP and flavor at high energy as well as at low energy (after spontaneous symmetry breaking, the low energy effective symmetries are still a combined theory of CP and flavor). Ref. \cite{cp12,cp24} combines a $U(1)$ global symmetry and its discrete subgroups such as $\mathbb{Z}_8$ with $\mu\tau $ reflection to have texture zeros  in light neutrino mass matrices so that the model could predict neutrino parameters in both the sectors, masses as well as mixing. Due to the blindness in the mixing sector, a combination of $\mu\tau$ symmetry with FL symmetry has been proposed in \cite{fl7}. Similar to these models, in our work, FL symmetry could be thought of as a complementary symmetry to the generalized $\mu\tau$ reflection and vice versa, rather than treating any of them (FL or general $\mu\tau$) as an expedient partner of each other.\\

Amongst many of the interesting results (which we shall discuss in the next section) that emerge as a consequence of the transformation in \eqref{fl}, it is worthwhile to stress two important departures from ${\rm CP}^{\mu\tau}$.

\hspace{1mm}$\bullet$ First of all, as mentioned earlier, $G_{lm}^{\mu\tau\theta}$ in \eqref{a2} is  a $\mu\tau$ mixing symmetry. It reduces to `$\mu\tau$-interchange' in the limit $\theta\rightarrow \pi/2$ which we address in rest of this paper as  `$\mu\tau$-interchange limit (MTIL)'. It is now trivial to anticipate that the mixing parameter $\theta (\neq\pi/2)$ conspires for the departure from maximal $\delta$ and $\theta_{23}$. However, we show in this paper that despite the generalization from ${\rm CP}^{\mu\tau}$ to ${\rm CP}^{\mu\tau\theta}$, the additionally imposed FL symmetry only allows a tiny deviation from the maximality of $\delta$ in this model.

\hspace{1mm}$\bullet$ The first condition in \eqref{z1} is satisfied for a nontrivial eigenvector $\boldsymbol{\eta}$ if $\det M_\nu=0$ which means at least one of the light neutrino masses is zero. Thus, by construction, this model predicts the absolute light neutrino mass scale.

For a consistent phenomenological analysis, apart from fitting the neutrino oscillation global-fit data, we study here the impact of ${\rm CP}^{\mu\tau\theta}$ symmetry on $\nu_\mu\rightarrow \nu_e$ oscillation  in the long baseline experiments such as NO$\nu$A, T2K and DUNE. In addition, in the context of recent discovery of high energy neutrino events at IceCube\cite{ice1,ice2,ice3,ice4,ice5}, assuming high energy neutrinos originate purely from distant astrophysical sources\footnote{We consider high energy neutrinos originating from $pp$ and $p\gamma$ collisions.}, we also calculate the flux-ratios which will be measured with enhanced statistics at advanced neutrino telescopes (e.g. IceCube and ANTARES\cite{antares}) in near future. These calculations show that any potential deviation from the democratic 1:1:1 distribution of flux ratios\cite{dem,dem1,dem2,dem3} can lead to predictions on the octant of $\theta_{23}$ in our model.


The rest of the paper is organized as follows. Sec.\ref{sec2} contains the most general parametrization of $M_\nu$ that is invariant under \eqref{fl}, thereby satisfying the conditions of \eqref{z1}. Sec.\ref{sec3} deals with the evaluation of Majorana phases $\alpha,\beta$ and the leptonic Dirac CP phase $\delta$ for both types of mass ordering analysed in two different subsections. The numerical analysis in Sec.\ref{sec4} comprises of four subsections. Subsec.\ref{sec4a} entails the extraction of the allowed parameter space and the prediction of light neutrino masses, whereas Subsec.\ref{sec4b} deals with the prediction on neutrinoless double beta decay process. Subsec.\ref{sec4c} discusses of the range of variation of the oscillation probability $P_{\mu e}$ and the CP asymmetry parameter $A_{\mu e}$ in experiments such as T2K, NO$\nu$A and DUNE for both NO and IO. Subsec.\ref{sec4d} comments on the possibility of determining the octant of $\theta_{23}$ from futuristic measurements of flavor flux ratios in neutrino telescopes such as IceCube.

\section{FL transformed  ${\rm CP}^{\mu\tau\theta}$ invariance of $M_\nu$}\label{sec2}
  Using \eqref{z1},   a $3\times 3$ symmetric mass matrix can   most generally be parametrized as\footnote{In rest of the paper, $\eta _e$, $\eta _{\mu}$ and $\eta _{\tau}$ are referred to as $\eta _1$, $\eta _2$ and $\eta _3$ respectively.}:
\begin{equation}
M_\nu=\\ \begin{pmatrix}
-\frac{2a_1}{(1+c_\theta)}\frac{\eta _2}{\eta _1} & a_1+ia_2 & -a_1t_{\frac{\theta}{2}}+ia_2t^{-1}_{\frac{\theta}{2}}\\
a_1+ia_2 & c_1t_{\frac{\theta}{2}}-a_1\frac{\eta _1}{\eta _2}-ia_2(1+c_\theta)\frac{\eta _1}{\eta _2} & c_1-ia_2t^{-1}_{\frac{\theta}{2}}c_{\theta}\frac{\eta _1}{\eta _2}\\
-a_1t_{\frac{\theta}{2}}+ia_2t^{-1}_{\frac{\theta}{2}} & c_1-ia_2t^{-1}_{\frac{\theta}{2}}c_{\theta}\frac{\eta _1}{\eta _2} & c_1t^{-1}_{\frac{\theta}{2}}-a_1\frac{\eta _1}{\eta _2}+ia_2(1+c_\theta)\frac{\eta _1}{\eta _2}
\end{pmatrix},
\label{massCPmutauFL}
\end{equation}
where $c_\theta\equiv\cos\theta, s_\theta\equiv \sin\theta$ and $t_{\theta/2}=\tan\frac{\theta}{2}$. For simplicity, we restrict to a reasonable choice that $\eta_l$ are a priori arbitrary complex numbers with same phases, so that the ratios $\frac{\eta_1}{\eta_1},\frac{\eta_2}{\eta_3}$ and $\frac{\eta_3}{\eta_1}$ are all real.
In \eqref{massCPmutauFL}, there are five real free parameters: $a_1$, $a_2$, $c_1$, $\frac{\eta _1}{\eta _2}$ and $\theta$ which can be well constrained by existing neutrino oscillation global-fit data. It is to be noted that~\eqref{massCPmutauFL} does not contain the parameter $\eta_3$ owing to a consistency relation of the form $\frac{\eta_2}{\eta_3}=-\frac{(1+c_\theta)}{s_\theta}$. The mass matrix $M_\nu$ in \eqref{massCPmutauFL}  can be diagonalized by a similarity transformation with a unitary matrix $U$:
\begin{equation}U^TM_\nu U=M_\nu^d \equiv {\rm diag}\hspace{1mm}(m_1,m_2,m_3), \label{e0}\end{equation}
 where ${m}_i\hspace{1mm}(i=1,2,3)$ are real and we assume that $m_i\geq 0$.
Without any loss of generality, we work in the diagonal basis of the charged lepton so that $U$ can be related to the $PMNS$ mixing matrix $U_{PMNS}$ as
{\small{
\begin{equation}
U=P_\phi U_{PMNS}\equiv
P_\phi \begin{pmatrix}
c_{1 2}c_{1 3} & e^{i\frac{\alpha}{2}} s_{1 2}c_{1 3} & s_{1 3}e^{-i(\delta - \frac{\beta}{2})}\\
-s_{1 2}c_{2 3}-c_{1 2}s_{2 3}s_{1 3} e^{i\delta }& e^{i\frac{\alpha}{2}} (c_{1 2}c_{2 3}-s_{1 2}s_{1 3} s_{2 3} e^{i\delta}) & c_{1 3}s_{2 3}e^{i\frac{\beta}{2}} \\
s_{1 2}s_{2 3}-c_{1 2}s_{1 3}c_{2 3}e^{i\delta} & e^{i\frac{\alpha}{2}} (-c_{1 2}s_{2 3}-s_{1 2}s_{1 3}c_{2 3}e^{i\delta}) & c_{1 3}c_{2 3}e^{i\frac{\beta}{2}}
\end{pmatrix},\label{eu}
\end{equation}}}
where $P_\phi={\rm diag}~(e^{i\phi_1},~e^{i\phi_2},~e^{i\phi_3})$ is an unphysical diagonal  phase matrix and  $c_{ij}\equiv\cos\theta_{ij}$, $s_{ij}\equiv\sin\theta_{ij}$ with the mixing angles $\theta_{ij}\in[0,\pi/2]$. We work within the PDG convention\cite{Tanabashi:2018oca} but denote our Majorana phases by $\alpha$ and $\beta$. CP-violation enters through nontrivial values of the Dirac phase $\delta$ and of the Majorana phases $\alpha,\beta$  where $\delta,\alpha,\beta\in[0,2\pi]$.

\section{Impact of mass ordering on mixing angles and CP properties }
\label{sec3}
Eqs.\eqref{z1} and \eqref{e0} jointly imply\cite{cp4}
\begin{equation} G^{\theta}U^*=U\tilde{d}. \label{atto}\end{equation} where $\tilde{d}={\rm diag}(\tilde{d}_1,\tilde{d}_2,\tilde{d}_3)$, where each $\tilde{d}_i$ $(i = 1, 2, 3)$ is either $+1$ or $-1$, and therefore \eqref{atto} can be written in the following explicit form:
\begin{equation}
\begin{pmatrix}-1 & 0 & 0\\0 & -c_\theta & s_\theta\\
0 & s_\theta & c_\theta\end{pmatrix}\begin{pmatrix}
U^*_{e1} & U^*_{e2} & U^*_{e 3}\\U^*_{\mu 1} & U^*_{\mu2} & U^*_{\mu3}\\
U^*_{\tau1} & U^*_{\tau2} & U^*_{\tau3}
\end{pmatrix}=\begin{pmatrix}
\tilde{d}_1 U_{e1} & \tilde{d}_2 U_{e2} & \tilde{d}_3 U_{e 3}\\
\tilde{d}_1 U_{\mu 1} & \tilde{d}_2 U_{\mu2} & \tilde{d}_3 U_{\mu3}\\
\tilde{d}_1 U_{\tau1} & \tilde{d}_2 U_{\tau2} & \tilde{d}_3 U_{\tau3}
\end{pmatrix}. \label{oju}
\end{equation}
Eq.\eqref{oju} is equivalent to nine equations for the three rows:
\begin{align}
-U^*_{e1}=\tilde{d}_1U_{e1}{\rm ,}~~~~~~~~~~-U^*_{e2}=\tilde{d}_2U_{e2},~~~~~~~~~~-U^*_{e2}=\tilde{d}_3U_{e3},&& \nonumber \\
-U^*_{\mu 1}c_\theta + U^*_{\tau 1}s_\theta = \tilde{d}_1U_{\mu 1}{\rm ,}-U^*_{\mu 2}c_\theta+U^*_{\tau 2}s_\theta = \tilde{d}_2U_{\mu 2},~-U^*_{\mu 3}c_\theta+U^*_{\tau 3}s_\theta =\tilde{d}_3U_{\mu 3} \nonumber \\
U^*_{\mu 1}s_\theta+U^*_{\tau 1}c_\theta=\tilde{d}_1U_{\tau 1},~U^*_{\mu 2}s_\theta+U^*_{\tau 2}c_\theta=\tilde{d}_2U_{\tau 2},~U^*_{\mu 3}s_\theta+U^*_{\tau 3}c_\theta=\tilde{d}_3U_{\tau 3} \label{relo}
\end{align}
It is useful to construct the following two rephasing invariant quantities, that are independent of the unphysical phases, for calculating the Majorana phases:
\begin{equation}
I_1=U_{e1}U^*_{e2}, ~~~~ I_2=U_{e1}U^*_{e3}.\label{3}
\end{equation}
From the first row of \eqref{relo}, we get,
\begin{equation}
I_1=\tilde{d}_1\tilde{d}_2U^*_{e1}U_{e2}, ~~~~ I_2=\tilde{d}_1\tilde{d}_2U^*_{e1}U_{e3} \label{4}
\end{equation}
Again, using the above different expressions for $I_{1,2}$, in \eqref{3} and \eqref{4}, we find the following relations,
\begin{equation}
c_{12}s_{12}c^2_{13}e^{-i\alpha/2}=\tilde{d}_1\tilde{d}_2 c_{12}s_{12}c^2_{13}e^{i\alpha/2} \label{5}
\end{equation}
and
\begin{equation}
c_{12}s_{13}c_{13}e^{i(\delta-\beta/2)}=\tilde{d}_1\tilde{d}_3 c_{12}s_{13}c_{13}e^{-i(\delta-\beta/2)}\label{6}.
\end{equation}
From \eqref{5} and \eqref{6}, we find,
\begin{equation}
e^{-i\alpha}=\tilde{d}_1\tilde{d}_2,~ e^{2i(\delta-\beta/2)}=\tilde{d}_1\tilde{d}_3,\label{p0}
\end{equation}
i.e., either $\alpha=0$ or $\alpha=\pi$, and either $\beta=2\delta$ or $\beta=2\delta-\pi$ . Therefore, there are four possible distinct pairs of values for the Majorana phases.
From the third row of \eqref{relo}, taking the absolute square, we obtain,
\begin{align}
|U_{\tau 3}|^2&=(U^*_{\mu 3}s_\theta+U^*_{\tau 3}c_\theta)(U_{\mu 3}s_\theta+U_{\tau 3}c_\theta) \\
\Rightarrow \cot2\theta_{23}&=\cot\theta\cos(\phi_2-\phi_3).\label{p1}
\end{align}
 Similarly, the absolute square of the second relation in the third row in \eqref{relo} is devoid of the unphysical phase difference $(\phi_2-\phi_3)$, and we get,
 \begin{equation}
\cos^2\delta=\cos^2\theta\sin^2(\phi_2-\phi_3)=\frac{\cos^2\theta\sin^22\theta_{23}-\sin^2\theta\cos^2
2\theta_{23}}{\sin^22\theta_{23}}.\label{p2}
\end{equation}
Note that, both the relations, i.e., \eqref{p1} and \eqref{p2} reduce to the co-bimaximal prediction of ${\rm CP}^{\mu\tau}$ in the MTIL, as expected. We also stress that the relations \eqref{p0}, \eqref{p1} and \eqref{p2} hold irrespective of the neutrino mass ordering.

Now, due to FL invariance, $M_\nu$ has a vanishing eigenvalue with   corresponding normalized eigenvector given by
 \begin{equation}
\textbf{v}=N^{-1}\begin{pmatrix}-\frac{\eta_1}{\eta_2}\cot\frac{\theta}{2}\\ -\cot\frac{\theta}{2}\\ 1\end{pmatrix}e^{i\gamma}, ~~{\rm with}~~N=\Bigg[\Big(1+\frac{\eta^2_1}{\eta^2_2}\Big)\cot^2\frac{\theta}{2}+1\Bigg]^{1/2},
\end{equation}
where $\gamma$ is an arbitrary phase signifying that the normalized eigenvector is unique up to an overall phase. If the zero eigenvalue is associated with $m_1=0$ ($m_3=0$), we discover  additional consequences for the normal (inverted) ordering.
\subsection{Normal ordering}\label{sec3b}
\begin{figure}[H]
\centering
\includegraphics[scale=.75]{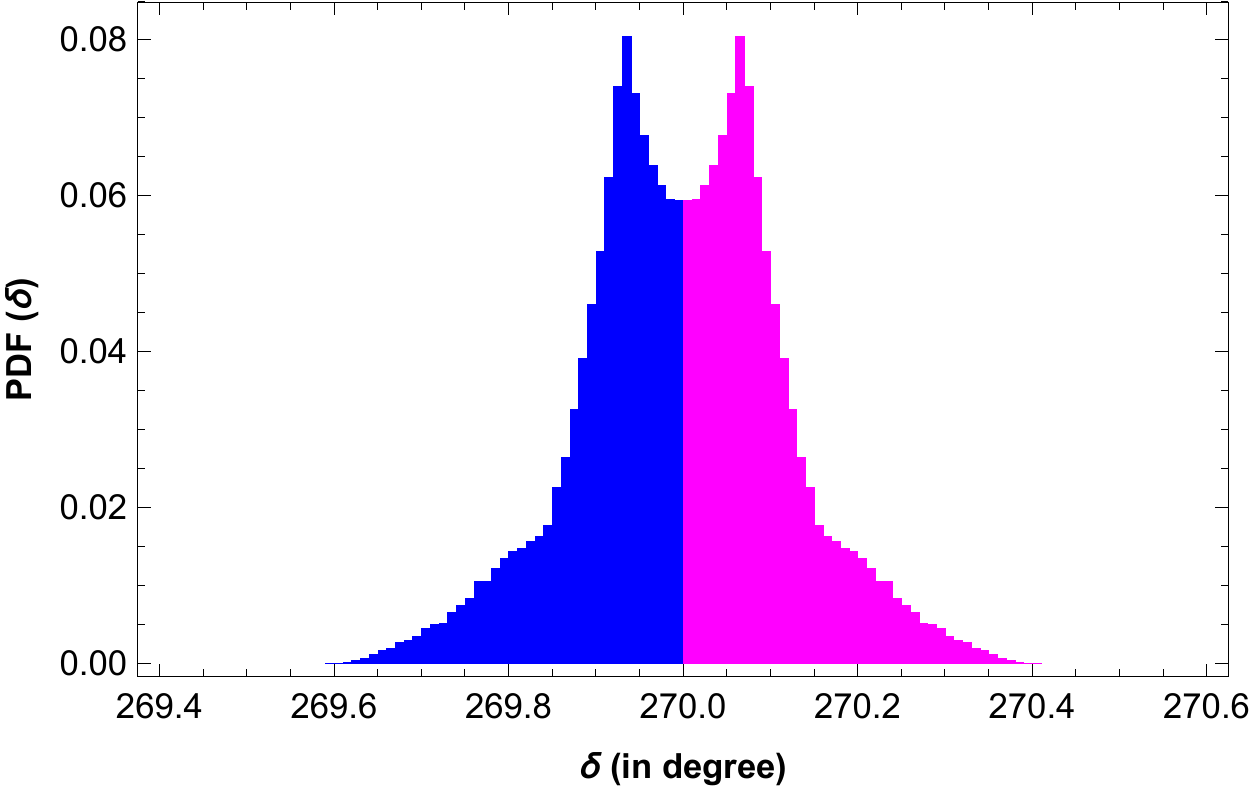}
\caption{Probability distribution of the Dirac CP phase $\delta$ for normal mass ordering. It is evident that the values which are very close to $270^\circ$ are most probable. To be  numerically  precise, $\int_{270}^{270\pm0.2}PDF(\delta) ~d\delta=0.795$. Thus upon a large number of random trial (we choose that number to be $10^6$), there is 80 $\%$ probability that $\delta$ will be in the range $270\pm 0.2$. }\label{fig1_prob}
\end{figure}

Here, $\textbf{v}$ is associated with the first column of PMNS.
Equating $\textbf{v}$ with the first column of $U$ in \eqref{eu}, we get, \begin{equation}c_{12}c_{13}=N^{-1}\frac{\eta_1}{\eta_2}\cot\frac{\theta}{2},~~\phi_1=\gamma+\pi,\label{r1}
\end{equation}
\begin{equation}
s_{1 2}c_{2 3}+c_{1 2}s_{2 3}s_{1 3} e^{i\delta}=N^{-1}\cot\frac{\theta}{2}e^{i(\gamma-\phi_2)},\label{r2}
\end{equation}
\begin{equation}
s_{1 2}s_{2 3}-c_{1 2}c_{2 3}s_{1 3} e^{i\delta}=N^{-1}e^{i(\gamma-\phi_3)}.\label{r3}
\end{equation}
Note that, \eqref{r2} and \eqref{r3} together imply \begin{equation} s^2_{12}=N^{-2}[\cot^2\frac{\theta}{2}+s^2_{23}+2s_{23}c_{23}\cot\frac{\theta}{2}\cos(\phi_2-\phi_3))].\label{r4}\end{equation}
Taking the product of \eqref{r2} with the complex conjugate of \eqref{r3}, and taking its imaginary part, we obtain,
\begin{equation}
\sin^2\delta=\frac{\cot^2\frac{\theta}{2}\sin^2(\phi_2-\phi_3)}{\Big[1+(1+\frac{\eta_1^2}{\eta_2^2})\cot^2\frac{\theta}{2}\Big]^2c^2_{12}s^2_{12}s^2_{13}}.
\end{equation} Eliminating $\sin^2(\phi_2-\phi_3)$ and using \eqref{p2}, we finally get


\begin{equation}
\cos^2\delta=\frac{\sin^22\theta_{12}s^2_{13}\cos^2\theta}{\sin^22\theta_{12}s^2_{13}\cos^2\theta+4\Big[1+(1+\frac{\eta_1^2}{\eta_2^2})\cot^2\frac{\theta}{2}\Big]^2
\cot^2\frac{\theta}{2}}. \label{nocosdel}
\end{equation}
Using \eqref{r4} and eliminating $\cos(\phi_2-\phi_3)$ from \eqref{p1}, we obtain, \begin{equation}
\cos^2\theta_{23}=\frac{\Big[\left\{1+(1+\frac{\eta_1^2}{\eta_2^2})\cot^2\frac{\theta}{2}\right \}s^2_{12}-1\Big]\cot\theta+\cot\frac{\theta}{2}}{(\cot^2\frac{\theta}{2}-1)\cot\theta+2\cot\frac{\theta}{2}}.\label{notheta23}
\end{equation}
 As we shall see in the numerical analysis in the next section, though in general $\cos\delta\neq 0$ for NO, the numerically allowed range of $\delta$ is very close to $3\pi/2$, lying in the narrow interval $269.6^\circ$ $-$ $270.4^\circ$ (Fig.~\ref{fig1_prob}). Since the possibility of $\delta=\pi/2$ is excluded at more than 99\% CL, by maximal CP violation, we refer only to $\delta = 3\pi /2$.

\noindent
 \subsection{Inverted ordering}\label{sec3a}
In this case, $\textbf{v}$ is associated with the third column of PMNS.
Equating $\textbf{v}$ with the third column of $U$ in \eqref{eu}, we get,
\begin{equation}
s_{13}=N^{-1}\frac{\eta_1}{\eta_2}\cot\frac{\theta}{2},~~\phi_1-\delta+\beta/2=\gamma+\pi,\label{s1}
\end{equation}
\begin{equation}
c_{13}s_{23}=N^{-1}\cot\frac{\theta}{2},~~\phi_2+\frac{\beta}{2}=\gamma+\pi,\label{s1m}
\end{equation}
\begin{equation}
c_{13}c_{23}=N^{-1},~~\phi_3+\frac{\beta}{2}=\gamma.\label{s2}
\end{equation}
Note that, \eqref{s1m} and \eqref{s2} together imply
\begin{equation}
\tan\theta_{23}=\cot\frac{\theta}{2},~~(\phi_2-\phi_3)=\pi,\label{iotheta23}
\end{equation}
 which is consistent with the relation \eqref{p1}. Note that, since the unphysical phase difference $(\phi_2-\phi_3)=\pi$, it follows from \eqref{p2} that the Dirac CP violation is maximal irrespective of the value of $\theta_{23}$ i.e.,
 \begin{equation}
\cos\delta=0.\label{iocosdel}
\end{equation}

Clearly, since the Dirac CP phase deviates slightly from its maximal value only for the NO, and both types of mass ordering in this model predict arbitrary nonmaximality in $\theta_{23}$, it is difficult to make comments on the mass ordering, only from the measurement of these two parameters. Though any large nonmaximality in $\delta$ will rule out ${\rm CP}^{\mu\tau}$ as well as this model (${\rm CP}^{\mu\tau\theta}+{\rm FL}$), however, if the experiments favour nonmaximal $\theta_{23}$ along with a maximal value of $\delta$ the latter model will survive while the former will be in tension.\\


One might wonder whether the minimal seesaw, which also leads to a vanishing eigenvalue, will lead to the same predictions as above when combined with general $\mu\tau$ symmetry. Though Eq.\ref{p2} holds for both the cases (combination of the generalized $\mu\tau$ reflection symmetry with minimal seesaw or FL symmetry), a closer inspection of Eq.~\ref{nocosdel} reveals in general predictions for $\cos\delta$ need not be the same. This is because in each case the model parameters are different and will be constrained differently by the neutrino oscillation data.

 \section{Numerical analysis}\label{sec4}
\subsection{Parameter Estimation}\label{sec4a}
We present a comprehensive numerical analysis to demonstrate the phenomenological viability of our proposal, and explore its implications on neutrino phenomenology in general. It is organized as follows. We utilize the ($3\sigma$) ranges of the globally fitted neutrino oscillation data\cite{nufit} together with the upper bound of $0.17~\rm eV$\cite{Aghanim:2016yuo} on the sum of the light neutrino masses from PLANCK and other cosmological observations in Table \ref{oscx}. The allowed range of parameters of $M_\nu$ are tabulated in Table \ref{osc2}. Subsequently, we discuss the predictions in our model on neutrinoless double beta decay, CP asymmetry in $\nu_\mu\to\nu_e$ oscillations and flavor flux ratios at neutrino telescopes in three separate subsections.
\begin{table}[H]
\begin{center}
\caption{Input values used in the analysis\cite{Esteban:2016qun}} \label{oscx}
 \begin{tabular}{|c|c|c|c|c|c|}
\hline
${\rm Parameter}$&$\theta_{12}$&$\theta_{23}$ &$\theta_{13}$ &$ \Delta
m_{21}^2$&$|\Delta m_{31}^2|$\\
&$\rm degrees$&$\rm degrees$ &$\rm degrees$ &$ 10^{-5}\rm
(eV)^2$&$10^{-3} \rm (eV^2)$\\
\hline
$3\sigma\hspace{1mm}{\rm
ranges\hspace{1mm}(NO)\hspace{1mm}}$&$31.42-36.05$&$40.3-51.5$&$8.09-8.98$&
$6.80-8.02$&$2.399-2.593$\\
\hline
$3\sigma\hspace{1mm}{\rm
ranges\hspace{1mm}(IO)\hspace{1mm}}$&$31.43-36.06$&$41.3-51.7$&$8.14-9.01$&
$6.80-8.02$&$2.369-2.562$\\
\hline
${\rm Best\hspace{1mm}{\rm fit\hspace{1mm}}values\hspace{1mm}(NO)}$ &
$33.62$ & $47.2$ &  $8.54$ &$7.40$ & $2.494$\\
\hline
${\rm Best\hspace{1mm}{\rm
fit\hspace{1mm}}values\hspace{1mm}(IO)}$&$33.62$&$48.1$&$8.58$&$7.40$&$2.465$\\
\hline
\end{tabular}
\end{center}
\end{table}
\noindent

\begin{table}[H]
\begin{center}
\caption{Output values of the parameters of $M_\nu$} \label{osc2}
 \begin{tabular}{|c|c|c|c|c|c|}
\hline
${\rm Parameters}$&$a_1/10^{-3}$&$a_2/10^{-3}$&$c/10^{-3}$&
$|\frac{\eta_1}{\eta_2}|$&$\theta^\circ$\\
\hline
${\rm NO}$& $-4.0-4.0$& $-6.5-6.5$& $-28-+28$&$+1.79-+2.11$&$79.6-101.6$\\
\hline
${\rm IO}$& $-2.7-+2.7$& $-36.0-+36.0$& $-11.6-+11.6$&$+0.18-+0.23$&$77.0-94.4$\\
\hline
\end{tabular}
\end{center}
\end{table}

\noindent
\subsection{Neutrinoless double beta ($0\nu\beta\beta$) decay process} \label{sec4b}
\begin{figure}[H]
\includegraphics[scale=.29]{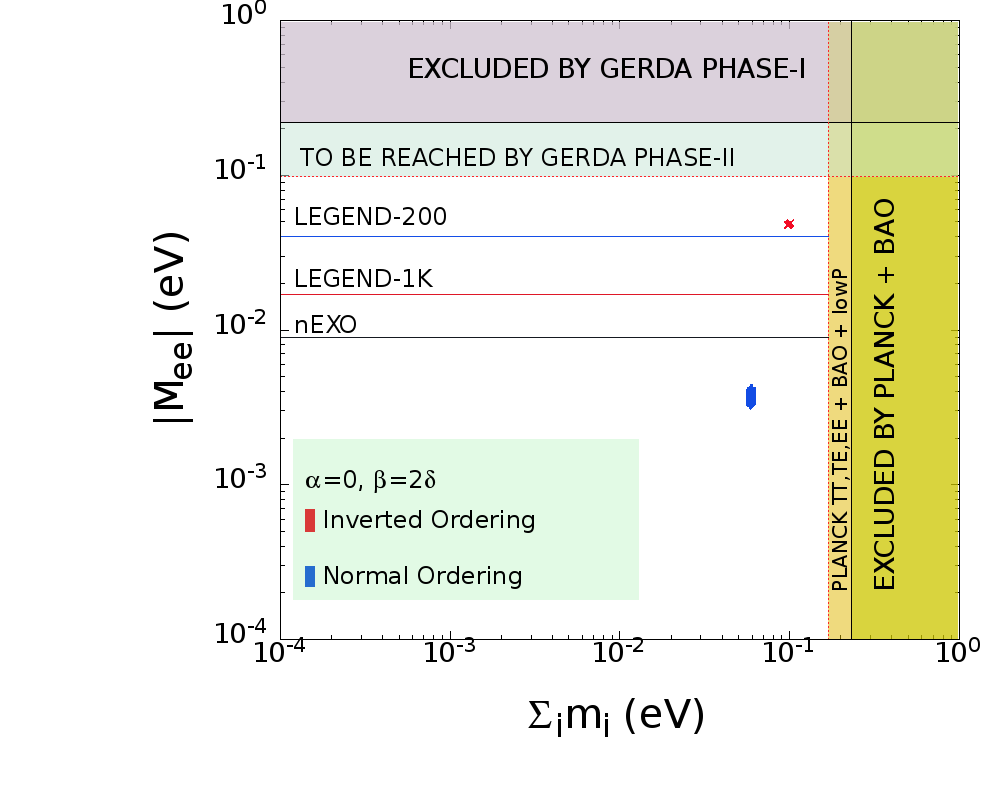}\includegraphics[scale=.29]{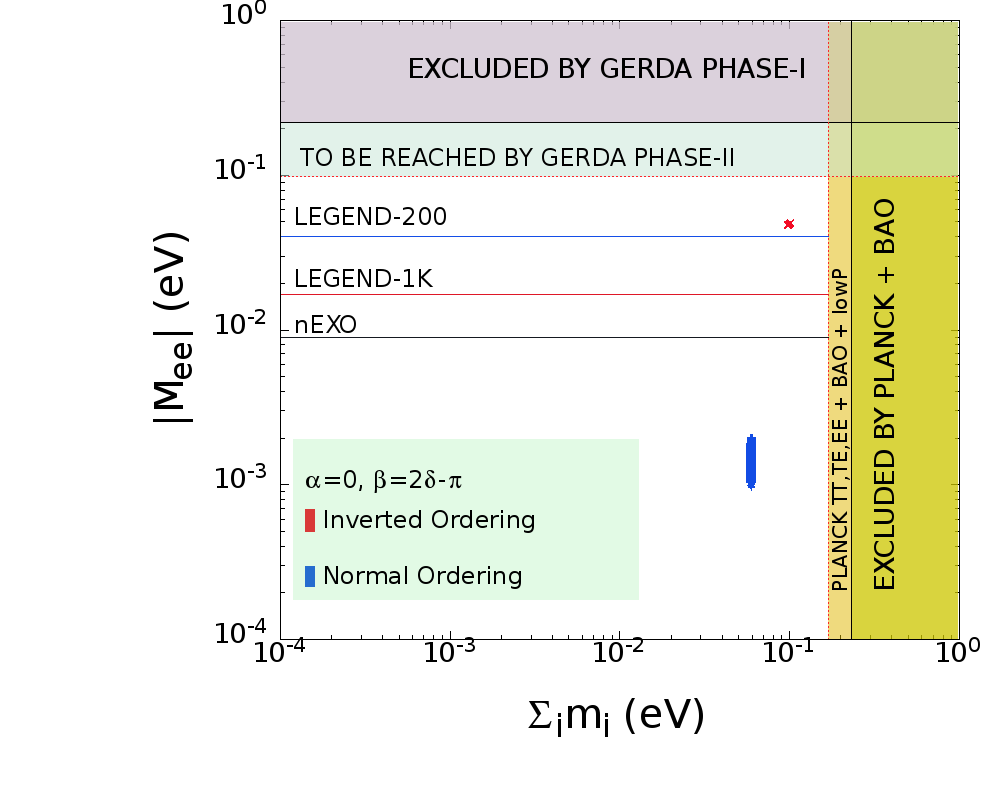}\\
\includegraphics[scale=.29]{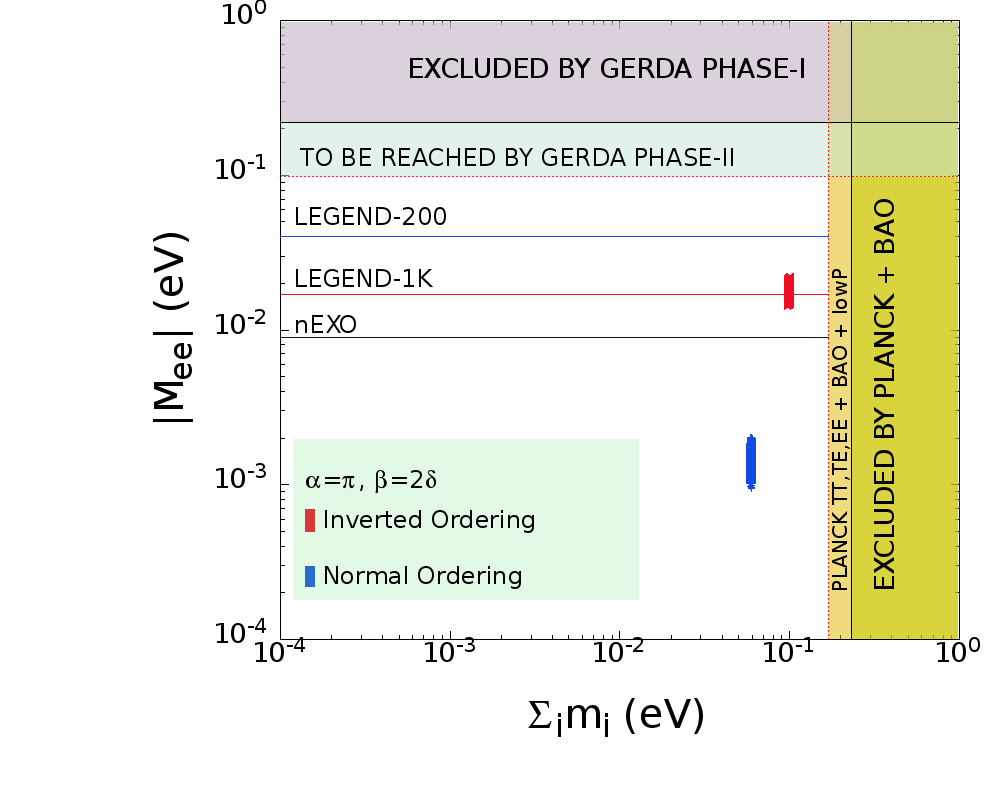}\includegraphics[scale=.29]{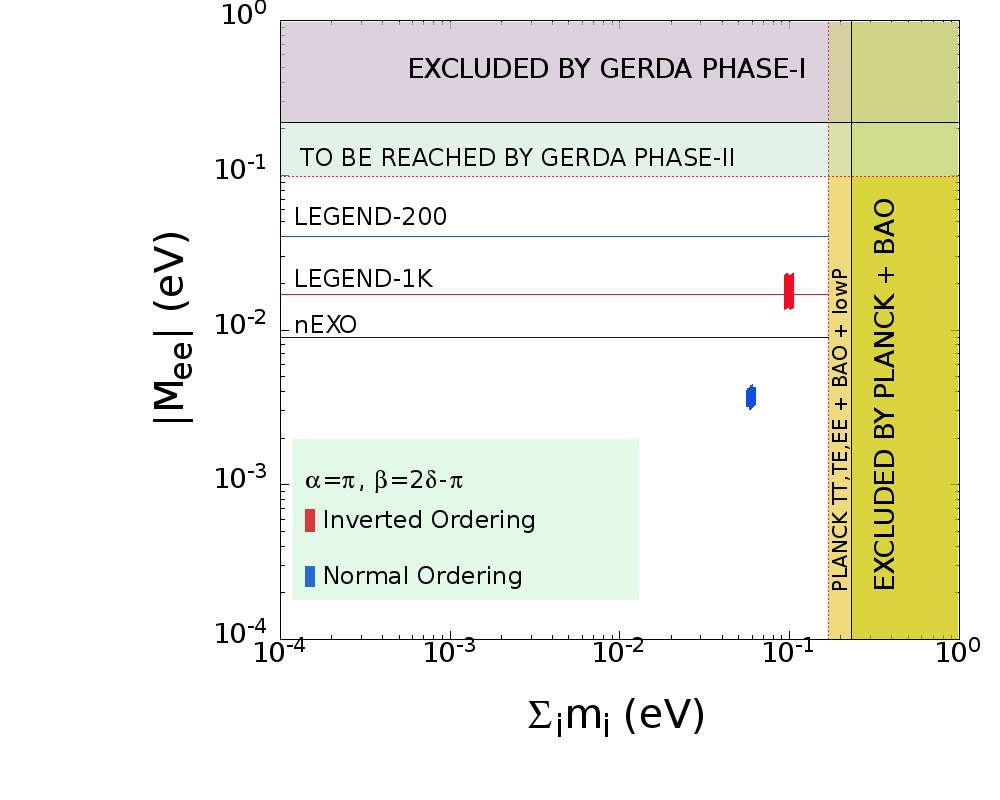}
\caption{Plots of $|M^{ee}|$ vs. $m_{min}$ for both types of mass ordering with four possible choices of the Majorana phases $\alpha$ and $\beta$.}\label{fg5}
\end{figure}

For certain nuclei such as Ge-76, it is energetically favorable to undergo a double beta decay ($2\nu\beta\beta$) instead of a singular $\beta-$decay emitting two electrons and two neutrinos. Moreover, if the neutrino is a Majorana particle those two neutrinos can annihilate each other to give rise to a neutrinoless double beta decay ($0\nu\beta\beta$): \begin{equation} (A,Z)\longrightarrow (A, Z+2)+2e^-
\end{equation} which clearly violates the lepton number by 2 units. Observation of such decay will firmly establish the Majorana nature of the neutrinos. The half-life corresponding to the above decay is given by
\begin{equation} \frac{1}{T^{0\nu}_{1/2}}=G_{0\nu}|\mathcal{M}|^2 |M_{ee}|^2m_e^{-2}, \end{equation} where $G_{0\nu}$ denote the two-body phase space factor, $\mathcal{M}$ is the nuclear matrix element (NME), $m_e$ is the mass of the electron and  $M_{ee}$ is the (1,1) element of the effective light neutrino mass matrix $M_\nu$. Using the PDG parametrization convention for $U_{PMNS}$, the $M_{ee}$ can be written as
\begin{equation}
M_{ee}=c_{12}^2c_{13}^2m_1+s_{12}^2c_{13}^2m_2e^{i\alpha}+
s_{13}^2m_3e^{i(\beta-2\delta)}.\label{alps}
\end{equation}
For the normal ordering, since $\delta$ deviates from $\pi/2$ or $3\pi/2$, and $m_1=0$ as a direct consequence of the FL symmetry, \eqref{alps} simplifies to the following four different possibilities for the four sets of $\alpha , \beta$ values as obtained in \eqref{p0} of Sec~\ref{sec3}:

(i) $\alpha=0,\beta=2\delta\Rightarrow M_{ee}=s^2_{12}c^2_{13}m_2+s^2_{13}m_3$,

(ii) $\alpha=0,\beta=2\delta-\pi\Rightarrow M_{ee}=s^2_{12}c^2_{13}m_2-s^2_{13}m_3$,

(iii) $\alpha=\pi,\beta=2\delta\Rightarrow M_{ee}=-s^2_{12}c^2_{13}m_2+s^2_{13}m_3$ and,

(iv) $\alpha=\pi,\beta=2\delta-\pi\Rightarrow M_{ee}=-s^2_{12}c^2_{13}m_2-s^2_{13}m_3$. Since the observations give upper bounds on $|M_{ee}|$, cases (i) and (iv) give identical predictions, as can be clearly seen from the upper left and lower right panels of Fig.\ref{fg5}. Similar situations occur for cases (ii) (upper right panel) and (iii) (lower left panel) in Fig.\ref{fg5}.
For the inverted ordering, $\delta= \pi/2$ or $3\pi/2$, and $m_3=0$. Here, due to the latter condition, the expression \eqref{alps} becomes independent of $\beta$ and reduces to two different possibilities:

(a) $\alpha=0,\beta=0,\pi\Rightarrow M_{ee}=c_{12}^2c_{13}^2m_1+s_{12}^2c_{13}^2m_2$,

(b) $\alpha=\pi,\beta=0,\pi\Rightarrow M_{ee}=c_{12}^2c_{13}^2m_1-s_{12}^2c_{13}^2m_2$.

 The plots of $|M_{ee}|$ versus the sum of the light neutrino masses $\sum\limits_{i}m_{i}$ for both NO and IO are displayed in Fig.\ref{fg5}. Several upper limits on $|M_{ee}|$ from various ongoing and upcoming experiments have been shown. It is evident from Fig.\ref{fg5} that  $|M_{ee}|$ in each plot leads to an upper limit which is below the sensitivity reach of the GERDA phase-II experimental data. The upper bounds on $|M^{ee}|$ from experiments such as LEGEND-200 ($40$ meV), LEGEND-1K ($17$ meV) and nEXO ($9$ meV)\cite{Agostini:2017jim}, shown in Fig.\ref{fg5}, can probe our model better. Note that, for each case, the entire parameter space corresponding to the inverted mass ordering is likely to be ruled out in case nEXO does not observe any $0\nu\beta\beta$ signal covering its entire reach.

Also the bounds on $\sum\limits_{i}m_{i}$ is projected to be improved in future cosmological observations. Upcoming Cosmic Microwave Background (CMB) experiments like CMB-S4 target the sensitivity $\sigma\sum\limits_{i}m_{i} \sim 20 ~\rm meV$ for a fiducial value of $\sum\limits_{i}m_{i} \simeq 58 ~\rm meV$.~\cite{Abazajian:2016yjj}. Future large scale structures observations in cosmology, such as galaxy surveys like DESI, Euclid, LSST~\cite{LSS-future} etc. are also projected to improve the bounds on $\sum\limits_{i}m_{i}$ while combined with the CMB observations~\cite{Lattanzi:2017ubx}. For example, a combination of WFIRST, Euclid, LSST and CMB Stage-III can achieve $\sigma\sum\limits_{i}m_{i} < 10 ~\rm meV$~\cite{LSS-comb}. These future bounds are particularly exciting in the predictions of $0\nu\beta\beta$ decays in neutrino mass models, as an upper bound of $\sum\limits_{i}m_{i} < 105 ~\rm meV$ will rule out IO.

\subsection{Effect of CP asymmetry in neutrino oscillations} \label{sec4c}
\begin{figure}[H]
\centering
\includegraphics[scale=0.6]{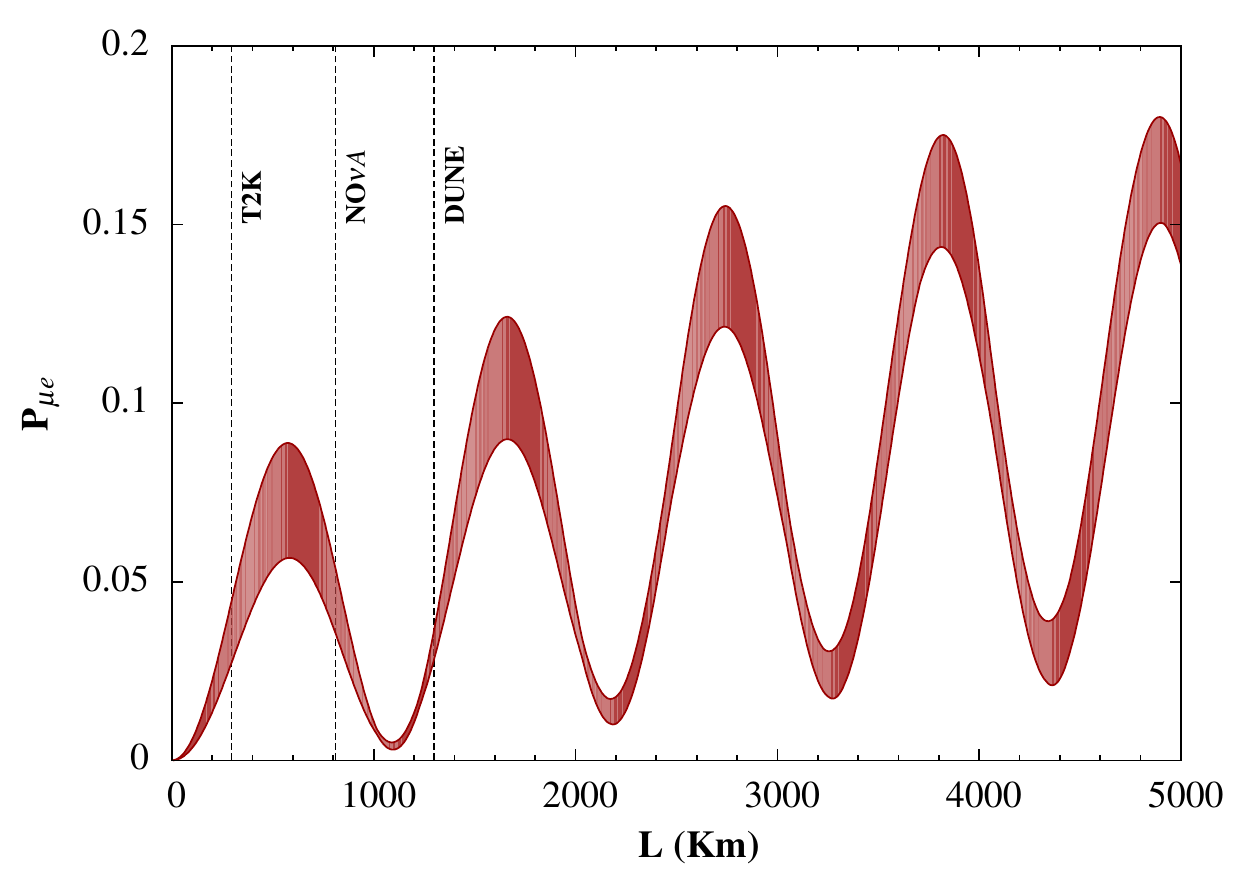}\includegraphics[scale=0.6]{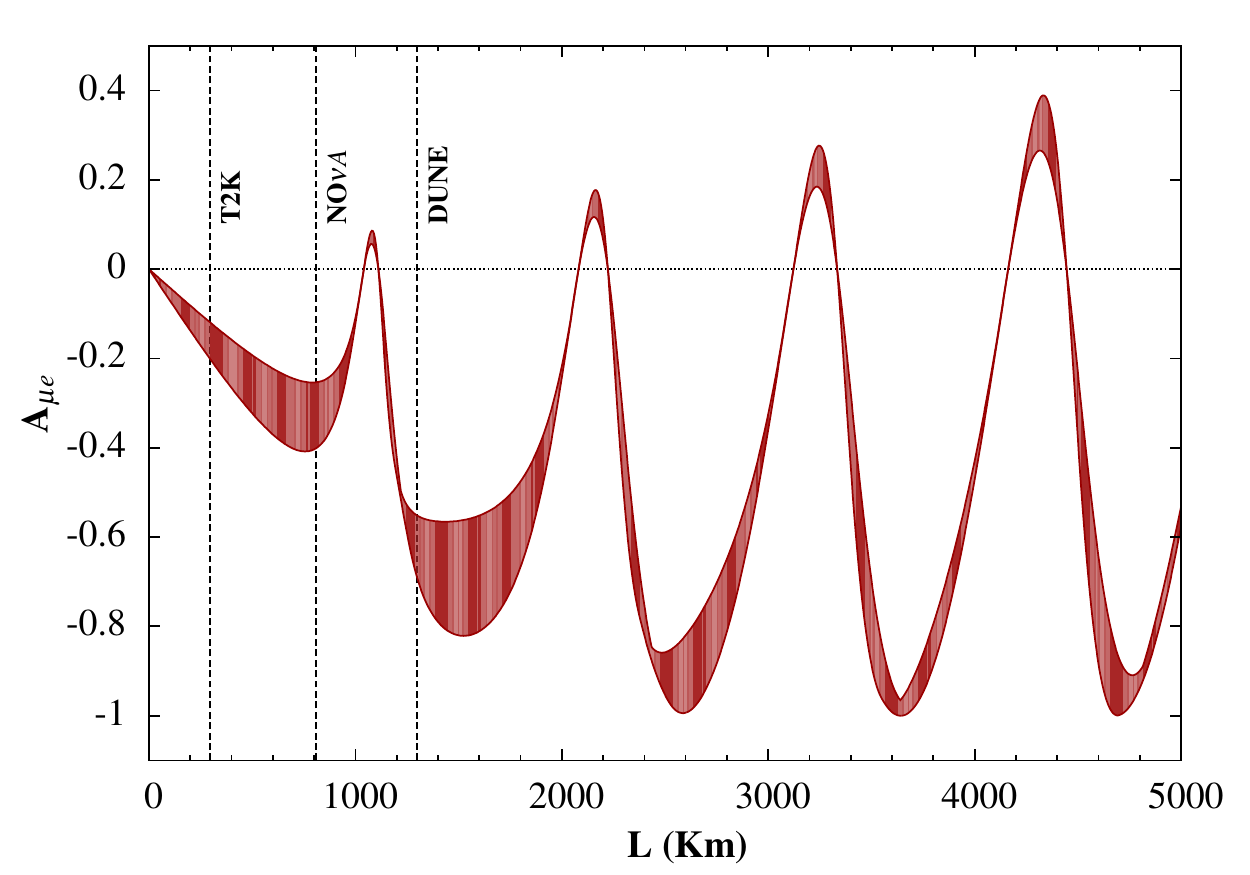}
\caption{Variation of the transition probability ($P_{\mu e}$) and CP asymmetry parameter ($A_{\mu e}$) against the baseline length $L$ for IO ($E = 1 ~\rm GeV$). The plots are for $\delta=3\pi /2$ and the bands correspond to $3\sigma$ ranges in $\theta_{12}$ and $\theta_{13}$. The three vertical dashed lines indicate observations at three different baseline lengths: $L=295 \rm Km$ for T2K, $L=810 \rm Km$ for No$\nu$A and $L=1300 \rm Km$ for DUNE. CP is conserved along the horizontal dotted line $A_{\mu e} = 0$.}\label{PmueAme_inv}
\end{figure}

\begin{figure}[H]
\centering
\includegraphics[scale=0.42]{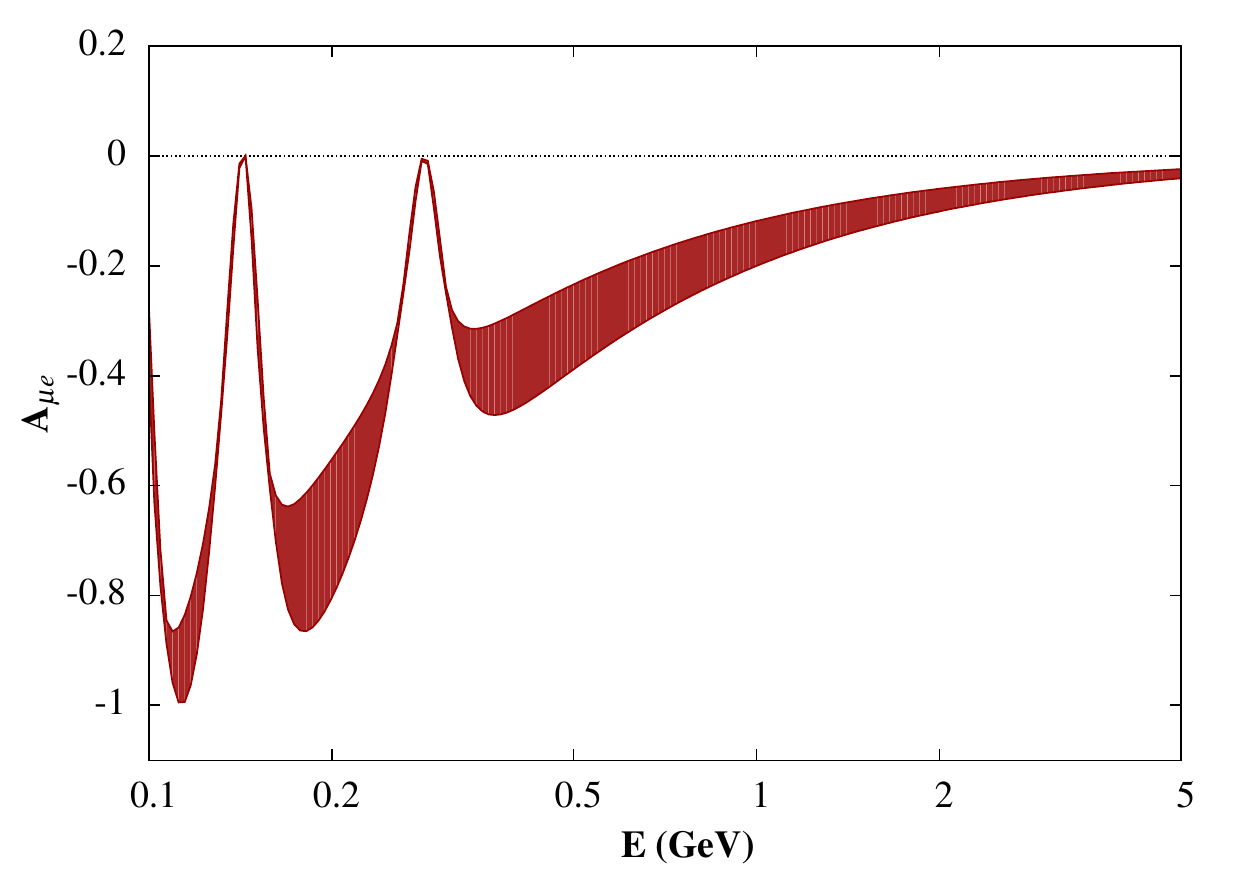}\includegraphics[scale=0.42]{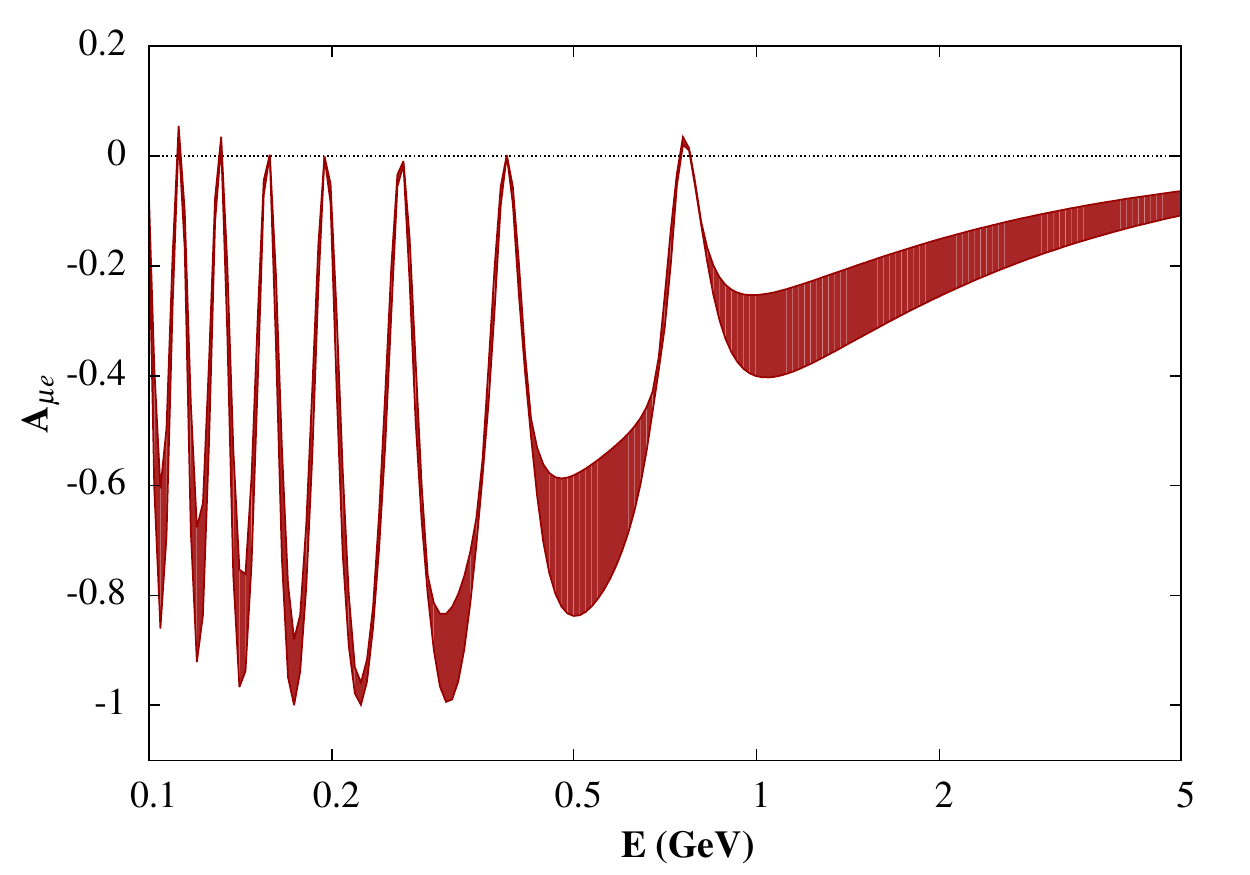}\includegraphics[scale=0.42]{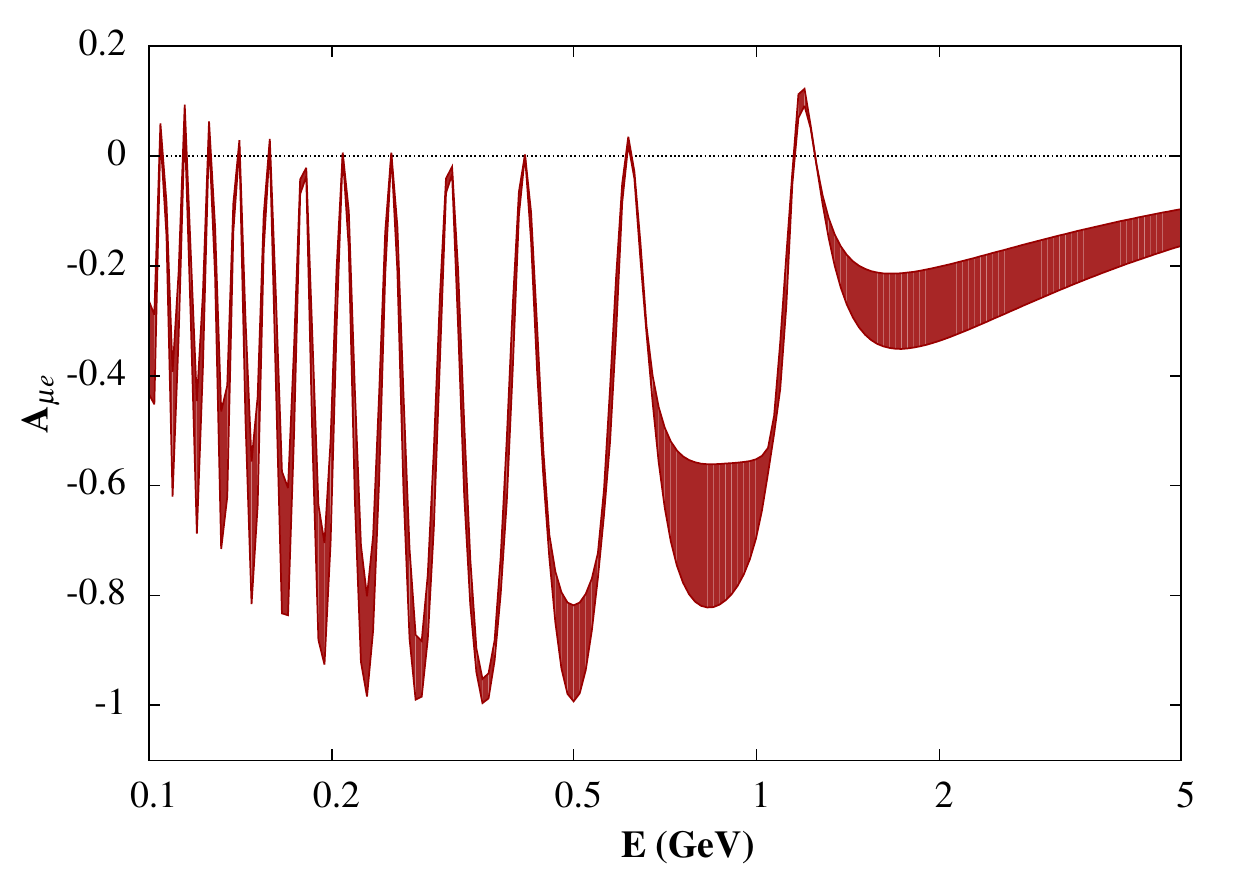}
\caption{Plots of the CP asymmetry parameter ($A_{\mu e}$) with energy E for fixed baseline lengths corresponding to different experiments in case of IO. Fig.(a) is for T2K with $L=295 \rm Km$; Fig.(b) is for NO$\nu$A with $L=810 \rm Km$ and Fig.(c) is for DUNE with $L=1300 \rm Km$. The plot is for $\delta=3\pi /2$, while the bands and the horizontal dashed lines have the same specifications as in Fig.~\ref{PmueAme_inv}.}\label{AmeE_inv}
\end{figure}

\begin{figure}[H]\centering
\includegraphics[scale=0.6]{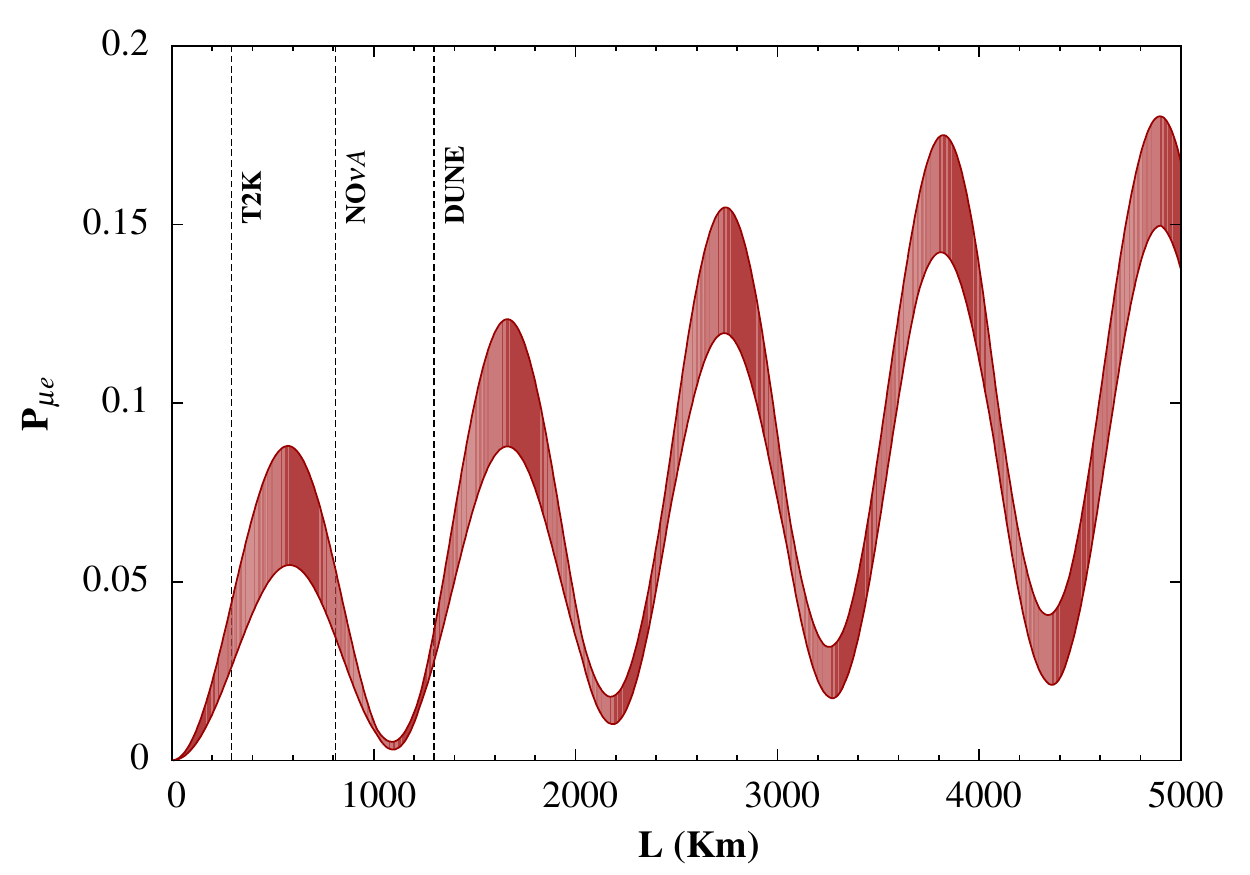}\includegraphics[scale=0.6]{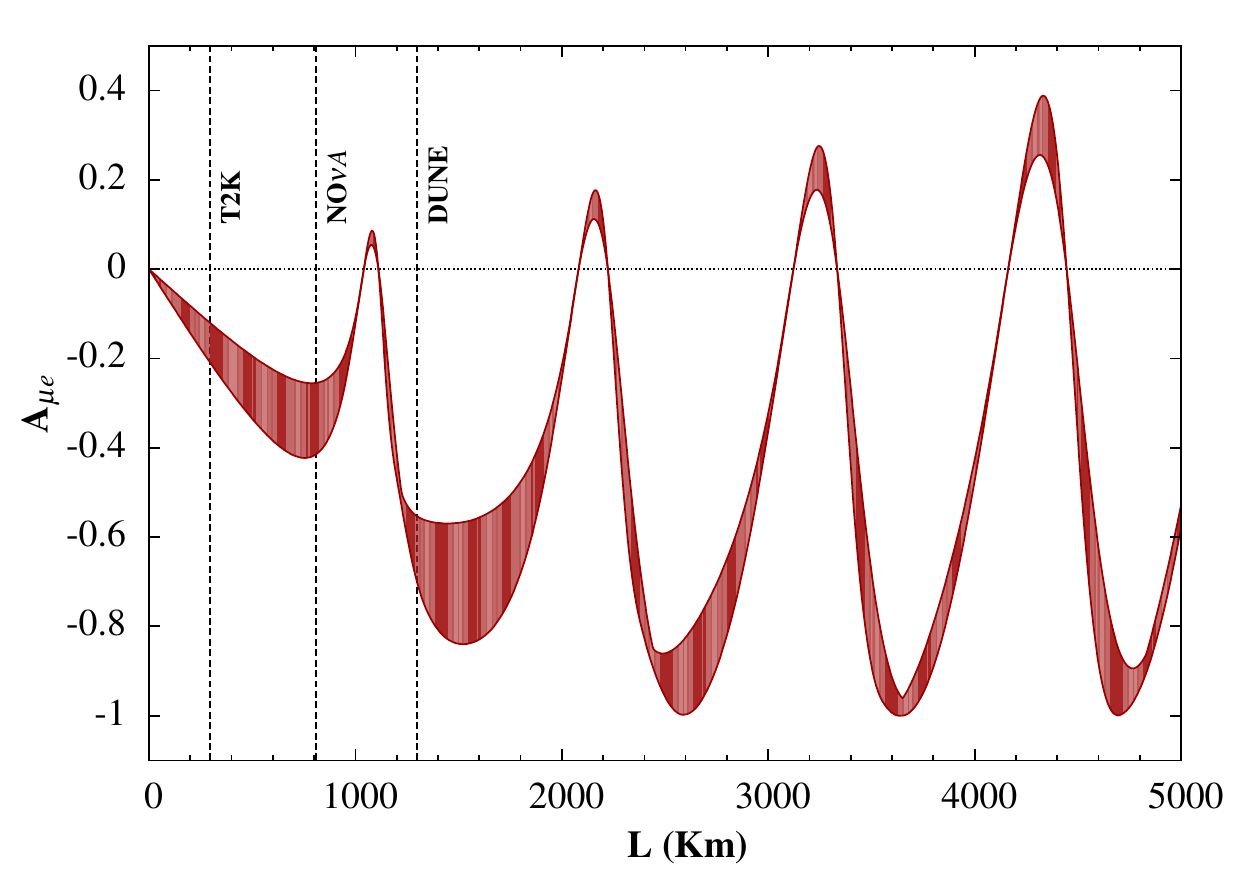}
\caption{Plots of the transition probability ($P_{\mu e}$) and CP asymmetry parameter ($A_{\mu e}$) with baseline length $L$ for NO ($E = 1 \rm GeV$). The bands are due to $3\sigma$ ranges of the mixing angles and also the ranges for the parameters $79.6^{\circ}<\theta < 101.6^{\circ}$ and $1.79<|\eta _1/\eta _2|<2.11$. In this case, $\delta$ is not fixed, but varies over a range predicted from \eqref{nocosdel} with the same ranges of the mixing angles, and model parameters $\theta$ and $\eta _1/\eta _2$. The three vertical dashed lines and the horizontal dotted line specify the same as Fig.\ref{PmueAme_inv}.}\label{PmueAme_nor}
\end{figure}

\begin{figure}[H]
\centering
\includegraphics[scale=0.42]{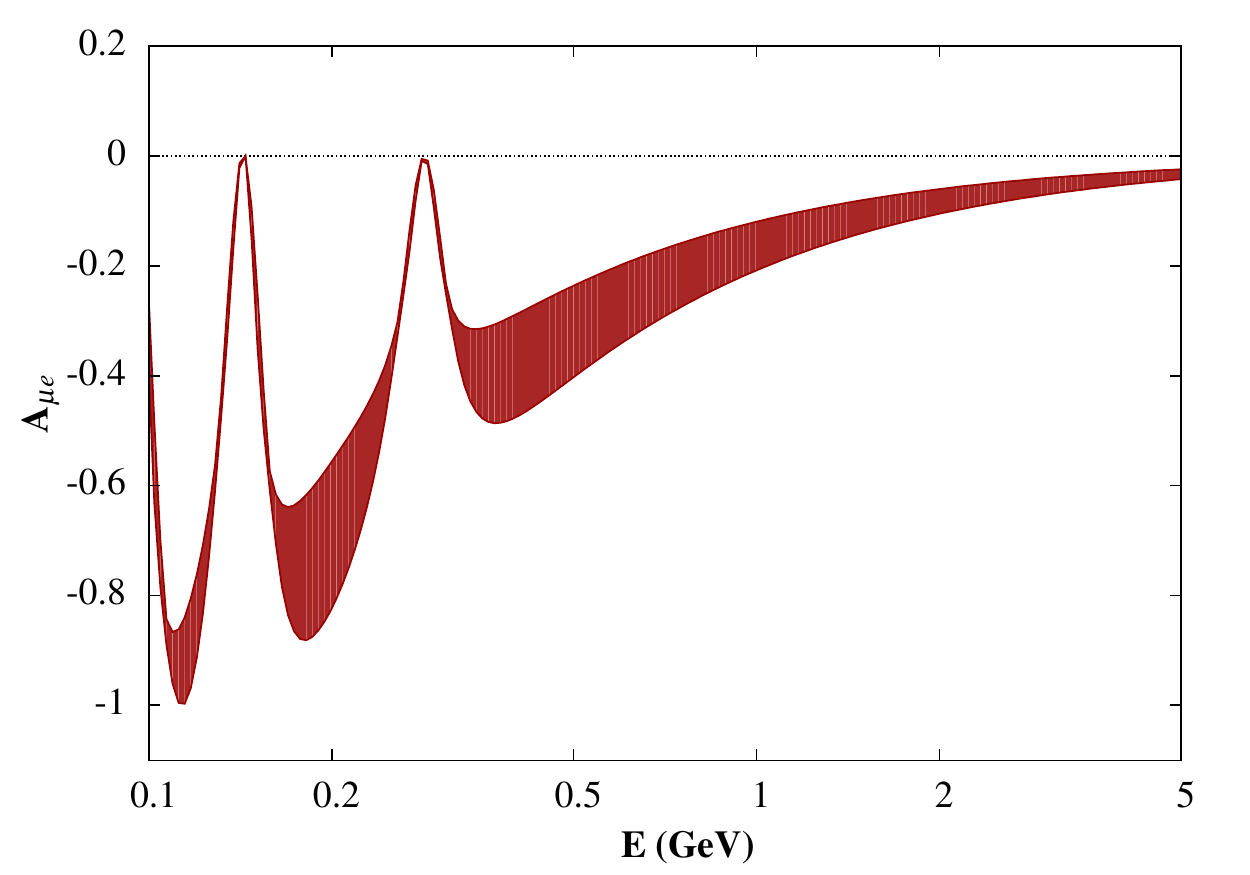}\includegraphics[scale=0.42]{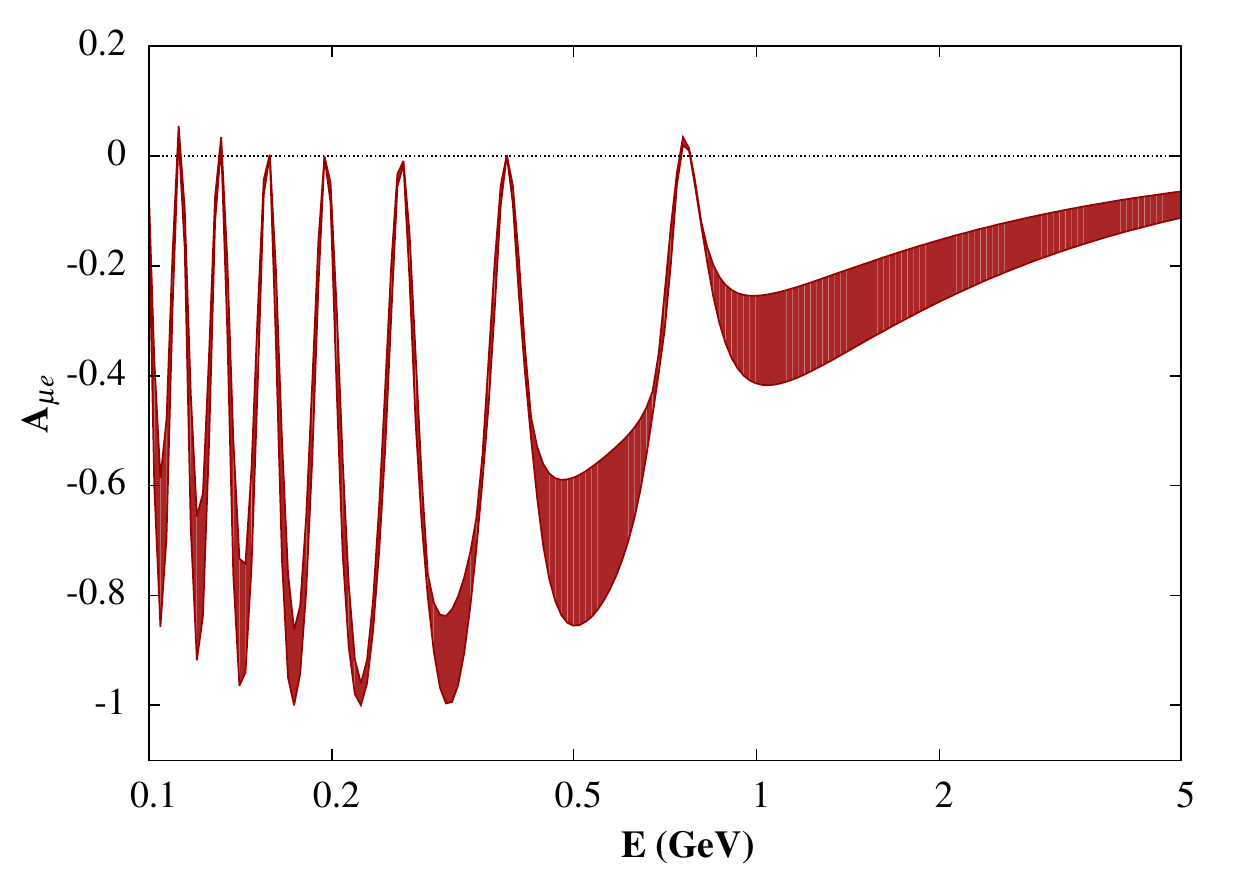}\includegraphics[scale=0.42]{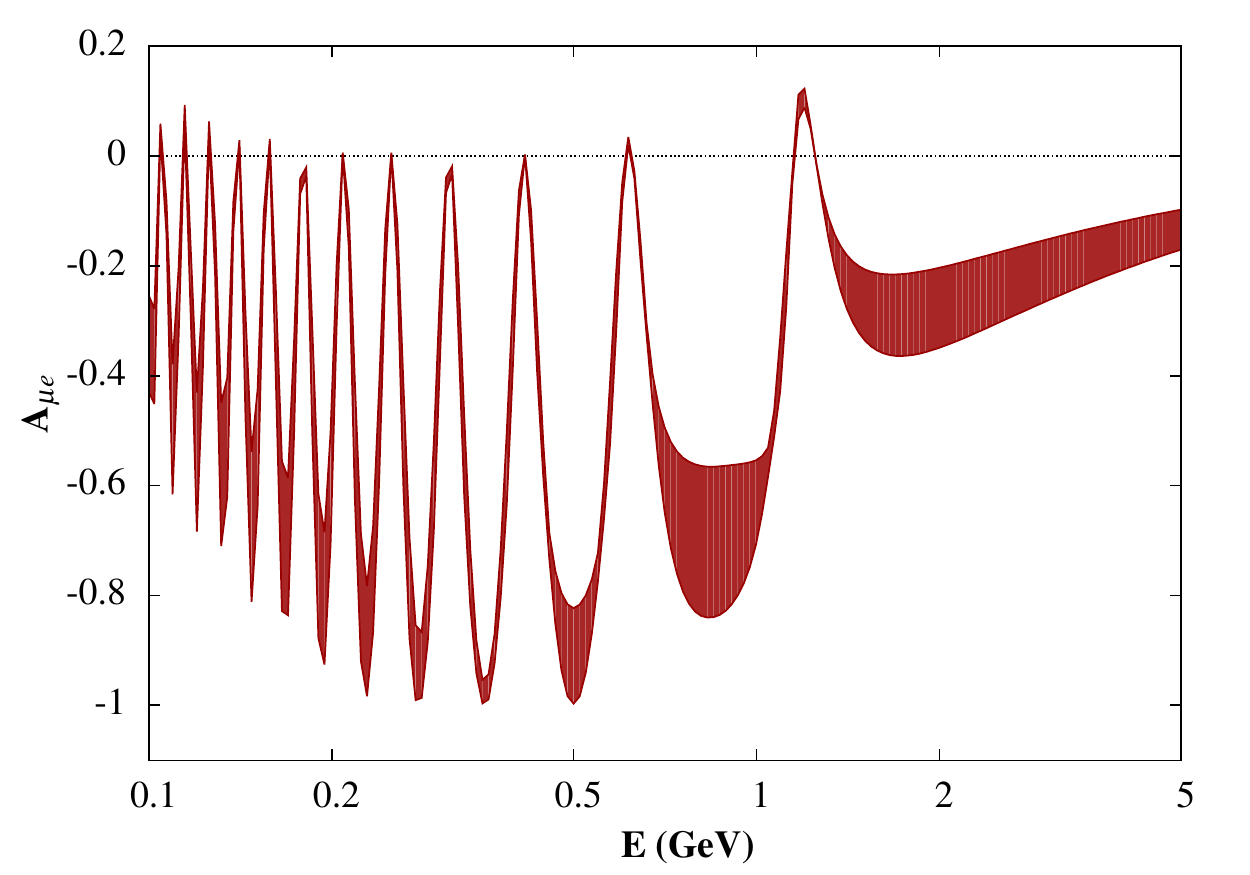}
\caption{Plots of the CP asymmetry parameter ($A_{\mu e}$) with energy $E$ for fixed baseline lengths corresponding to different experiments in case of NO. Fig.(a) is for T2K with $L=295 \rm Km$; Fig.(b) is for NO$\nu$A with $L=810 \rm Km$ and Fig.(c) is for DUNE with $L=1300 \rm Km$. The plots and their widths have same specifications as in Fig.~\ref{PmueAme_nor}. The horizontal lines denotes CP conservation ($A_{\mu e}=0$).}\label{AmeE_nor}
\end{figure}

In this section, we work out the effect of the presence of leptonic Dirac CP violation $\delta$ in neutrino oscillation experiments. The phase $\delta$ will appear in the asymmetry parameter $A_{lm}$, defined as \begin{equation}
A_{lm}=P(\nu_l\to\nu_m)-P(\bar{\nu}_l\to\bar{\nu}_m)
\end{equation} where $l,m=(e,\mu,\tau)$ are flavor indices and the $P$'s are transition probabilities. First, let us consider oscillation in vacuum. The $\nu_\mu\to \nu_e$ transition probability is given by
\begin{equation}
P_{\mu e}\equiv P(\nu_\mu\to \nu_e)=P_{atm}+P_{sol}+2\sqrt{P_{atm}}\sqrt{P_{sol}}\cos(\Delta_{32}+\delta),\label{x}
\end{equation}
where $\Delta_{ij}=\Delta m^2_{ij}L/4E$ is the kinematic phase factor ($L$ being the baseline length and $E$ being the beam energy) and $P_{atm},P_{sol}$ are respectively defined as
\begin{eqnarray} \sqrt{P_{atm}}=\sin\theta_{23}\sin\theta_{13}\frac{\sin(\Delta_{31}-aL)}{(\Delta_{31}-aL)}\Delta_{31},\\
\sqrt{P_{sol}}=\cos\theta_{23}\cos\theta_{13}\sin2\theta_{12}\frac{\sin aL}{aL}\sin\Delta_{21}.
\end{eqnarray}
Here $a=G_F N_e/\sqrt{2}$ with $G_F$ as the Fermi constant and $N_e$ is the number density of electrons in the medium of propagation, so that $a$ take into account the matter effects in neutrino propagation through the earth.  An approximate value of $a$ for the earth is $(3500{\rm km})^{-1}$\cite{Nunokawa:2007qh,Chen:2015siy}. In the limit $a\to 0$, \eqref{x} leads to the oscillation probability in vacuum. With this, the CP asymmetry parameter is given by \begin{equation}
A_{\mu e}=\frac{P(\nu_\mu\to\nu_e)-P(\bar{\nu}_\mu\to\bar{\nu}_e)}{P(\nu_\mu\to\nu_e)+P(\bar{\nu}_\mu\to\bar{\nu}_e)}=\frac{2\sqrt{P_{atm}}\sqrt{P_{sol}}\sin\Delta_{32}\sin\delta}{P_{atm}+2\sqrt{P_{atm}}\sqrt{P_{sol}}\cos\Delta_{32}\cos\delta+P_{sol}}
\end{equation} \
where $\delta$ is given by \eqref{nocosdel} and \eqref{iocosdel} for NO and IO respectively.
In Fig.\ref{PmueAme_inv} represents the variation of $P_{\mu e}$ and $A_{\mu e}$ against the baseline length $L$ for IO, i.e., for $\delta=3\pi/2$, while in Fig.\ref{PmueAme_nor} we give same plots for $\delta$ given by \eqref{nocosdel} i.e., for NO. The baseline lengths T2K, NO$\nu$A and DUNE are indicated in these figures by vertical lines. In Fig.\ref{AmeE_inv} and \ref{AmeE_nor} the CP asymmetry $A_{\mu e}$ is plotted against the beam energy $E$ for the same three experiments for IO and NO respectively.

\subsection{Octant of $\theta_{23}$ from flavor flux measurement at neutrino telescope} \label{sec4d}
Recent discovery\cite{ice1,ice2,ice3,ice4,ice5} of Ultra High Energy (UHE) neutrino events at IceCube has opened a new era in the neutrino astronomy. Including track+shower, IceCube has reported 82 high-energy starting events (HESE) which constitute more than 7$\sigma$ excess over the atmospheric background and thus points towards an extraterrestrial origin of the UHE neutrinos(for a recent update see Ref.\cite{ice6}). In addition, no significant spatial clustering has been found and the recent data seems to be consistent with isotropic neutrino flux from uniformly distributed point sources\cite{ice7} and hints towards extra galactic nature of the observed events. Although the HESE events are not consistent with the standard astrophysical one component unbroken isotropic power-law spectrum $\Phi (E_\nu)\propto E_\nu^{-2}$ and also suffer constraints from multi-messenger gamma-ray observation\cite{ice8}, two component  explanation of the observed neutrino flux from purely astrophysical sources is still a plausible scenario\cite{dev}. Before we discuss the predictions of our model based on the flavor flux ratios, statements on which could be made from enhanced statistics at neutrino telescopes (e.g., IceCube) and fits like\cite{dev}, we first lay out a short summary of the subject as a necessary prerequisite.

\begin{figure}[H]
\includegraphics[scale=0.75]{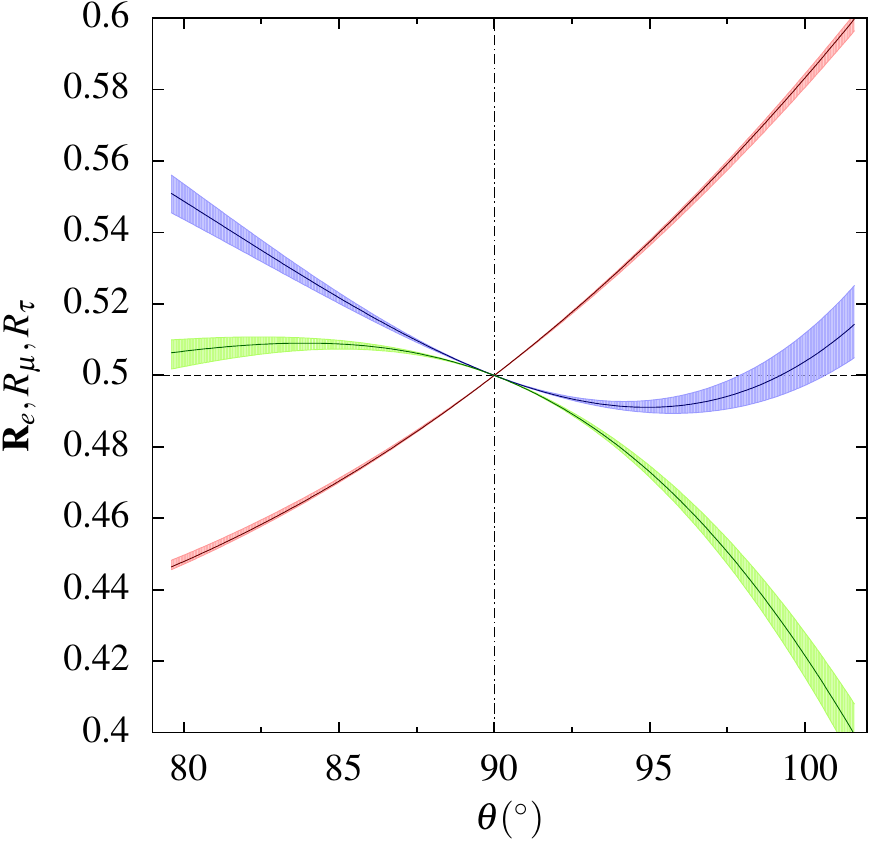}\hspace{0.5cm}\includegraphics[scale=0.75]{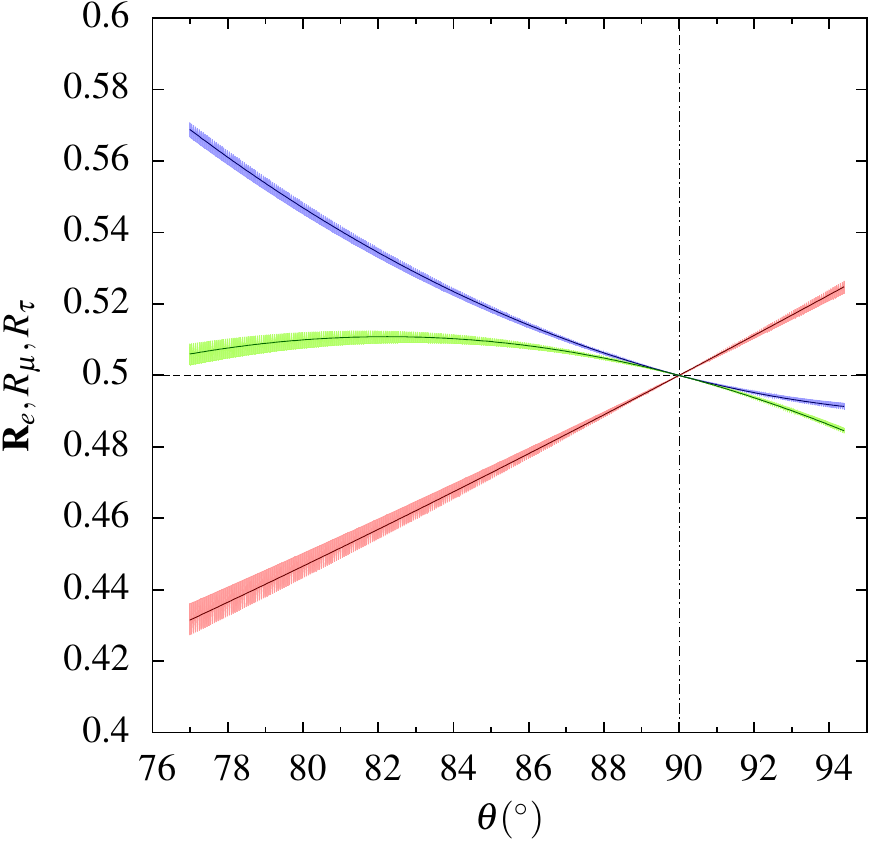}
\caption{Variation of the flavor the flux ratios $R_e$ (red), $R_{\mu}$ (blue) and $R_{\tau}$ (green) with $\theta$ for NO (left panel) and for IO (right panel). The solid lines represent plots for the best-fit values of the mixing angles and the bands are caused by the current 3$\sigma$ ranges of the mixing angles $\theta _{12}$ and $\theta _{13}$. The horizontal axes in both plots correspond to the numerically obtained ranges of $\theta$ in Table\ref{osc2}, which is different in NO and IO. For the NO case, $\eta _1 / \eta _2$ is fixed at $1.0$.}\label{ReRmuRtau}
\end{figure}

The dominant source of UHE cosmic neutrinos are $pp$ (hadro-nuclear) collisions in cosmic ray reservoirs such as galaxy clusters  and $p\gamma$ (photo-hadronic) collisions in cosmic ray accelerators\cite{ice9,ice10} such as gamma-ray bursts, active galactic nuclei and blazars. In  $pp$ collisions, protons of TeV$-$PeV range produce neutrinos via the decays $\pi^+\to \mu^+\nu_\mu, \pi^-\to \mu^-\bar{\nu}_\mu, \mu^{+}\to e^+\nu_e\bar{\nu}_\mu$ and $\mu^-\to e^{-}\bar{\nu}_e\nu_\mu.$ Therefore, the normalized flux distributions over flavor are\cite{dem2}
\begin{equation}
\{\phi^S_{\nu_e},\phi^S_{\bar{\nu}_e},\phi^S_{\nu_\mu},\phi^S_{\bar{\nu}_\mu},\phi^S_{\nu_\tau},\phi^S_{\bar{\nu}_\tau}\}=\phi_0\Big\{\frac{1}{6},\frac{1}{6},\frac{1}{3},\frac{1}{3},0,0\Big\},
\end{equation}
where the superscript $S$ denotes `source'. On the other hand, the $p\gamma$ collisions involve relatively less energetic $\gamma-$rays (GeV- $10^2$ GeV range). Therefore, the center-of-mass energy of $\gamma p$ system is such that it can only produce $\gamma p\to \Delta^+\to \pi^+ n$, which in turn give rise to decays $\pi^+\to \mu^+\nu_\mu$ and $\mu^+\to e^+\nu_e\bar{\nu}_\mu.$ The corresponding normalized flux distributions over flavor
\begin{equation}
\{\phi^S_{\nu_e},\phi^S_{\bar{\nu}_e},\phi^S_{\nu_\mu},\phi^S_{\bar{\nu}_\mu},\phi^S_{\nu_\tau},\phi^S_{\bar{\nu}_\tau}\}=\phi_0\Big\{\frac{1}{3},0,\frac{1}{3},\frac{1}{3},0,0\Big\}.
\end{equation}
In either case, if we take $\phi^S_l=\phi^S_{\nu_l}+\phi^S_{\bar{\nu}_l}$ with $l=e,\mu,\tau$, \begin{equation}
\{\phi_e^S,\phi_\mu^S,\phi_\tau^S\}=\phi_0\Big\{\frac{1}{3},\frac{2}{3},0\Big\}.
\end{equation}
As neutrino oscillations will change flavor distributions from source (S) to telescope (T)\cite{Xing:2012sj} the flux reaching the telescope will be given by
\begin{equation}
\phi^T_{l}=\phi^T_{\nu_l}+\phi^T_{\bar{\nu}_l}=\sum\limits_{m}
\Big[\phi^S_{\nu_m} P(\nu_m\to\nu_l)+\phi^S_{\bar{\nu}_m}P(\bar{\nu}_m\to
\bar{\nu}_l)\Big].
\end{equation}
Since the source-to-telescope distance is much greater than the oscillation length, the flavor oscillation probability averaged over many oscillations is given by
\begin{equation}
P(\nu_m\to\nu_l)=P(\bar{\nu}_m\to
\bar{\nu}_l) \approx \sum\limits_{i}|U_{l i}|^2|U_{m i}|^2.
\end{equation}
Thus the flux reaching the telescope is given by
\begin{equation}
\phi_l^T=\sum\limits_{i}\sum\limits_{m}\phi_m^S|U_{l i}|^2|U_{m i}|^2=\frac{\phi_0}{3}\sum\limits_{i}|U_{l i}|^2(|U_{ei}|^2+2|U_{\mu i}|^2)
\end{equation}
where $\phi_0$ is the overall flux normalization. The unitarity of the PMNS matrix implies \begin{equation}
\phi_l^T=\frac{\phi_0}{3}[1+\sum\limits_{i}|U_{l i}|^2(|U_{\mu i}|^2-|U_{\tau i}|^2)]=\frac{\phi_0}{3}[1+\sum\limits_{i}|U_{l i}|^2\Delta_i].
\end{equation}
where $\Delta_i=|U_{\mu i}|^2-|U_{\tau i}|^2$. Existence of exact $\mu\tau$ (anti)symmetry, therefore dictates that $\Delta_i=0$, and $\phi_e^T=\phi_\mu^T=\phi_\tau^T.$ With the above background, one can define certain flavor flux ratios $R_l$ ($l=e,\mu,\tau$) at the neutrino telescope as
\begin{equation}
R_l\equiv\frac{\phi_l^T}{\sum\limits_{m}\phi_m^T-\phi_l^T}=\frac{1+\sum\limits_{i}|U_{l i}|^2\Delta_i}{2-\sum\limits_{i}|U_{l i}|^2\Delta_i},\label{fluxratio}
\end{equation}
where $l,m=e,\mu,\tau$ and $U$ is given in \eqref{eu}. Each $R_l$ depends on all three mixing angles and $\cos\delta$. For NO, $\theta_{23}$ and $\cos\delta$ are given by \eqref{notheta23} and \eqref{nocosdel} while for IO the corresponding quantities are given by \eqref{iotheta23} and \eqref{iocosdel} respectively. For both types of ordering, we display in Fig.\ref{ReRmuRtau} the variation of $R_{e,\mu,\tau}$ w.r.t $\theta$ in its phenomenologically allowed ranges (Table \ref{osc2}) using the exact expressions in \eqref{fluxratio}.

For NO, $\theta_{23}$ can be eliminated in favor of $\theta$ and $\eta_1/\eta_2$. Keeping the latter fixed at a value $1.5$, we show in Fig.\ref{ReRmuRtau} (left panel) the contour corresponding to the best-fit values of $\theta_{12}$ and $\theta_{13}$, while the bands arise when $\theta_{12}$ and $\theta_{13}$ are allowed to vary in their current $3\sigma$ ranges. It should be emphasized that the contours corresponding to $\cos\delta>0$ and $\cos\delta<0$ are practically indistinguishable, and therefore, we show the contours and bands only for the case $\cos\delta>0$.

Next, in case of IO, $\theta_{23}$ can be eliminated in favor of $\theta$ only. The resulting variation of $R_{e,\mu,\tau}$ with $\theta$ are shown in the right panel of Fig.\ref{ReRmuRtau}. In generating these plots, the mixing angles $\theta _{12}$ and $\theta _{13}$ are again allowed to vary in their experimental $3\sigma$ ranges. The contours within the bands represent the case when $\theta_{12}$ and $\theta_{13}$ are kept fixed at their best-fit values. Unlike NO, the expressions for $R_l$ in case of IO are relatively simple and can be used to explain the nature of the plots. The expressions for $R_{e,\mu,\tau}$ for IO are:
%
%
%


\begin{eqnarray}
R_e&\approx&\frac{2-\sin^22\theta_{12}c_{\theta}}{4+\sin^22\theta_{12}c_{\theta}},\nonumber \\
R_\mu&\approx&\frac{1+\frac{1}{4}\sin^22\theta_{12}c_{\theta}+(1-\frac{1}{4}\sin^22\theta_{12})c_{\theta}^2}{2-\frac{1}{4}\sin^22\theta_{12}c_{\theta}-(1-\frac{1}{4}\sin^22\theta_{12})c_{\theta}^2},\nonumber \\
R_\tau&\approx&\frac{1+\frac{1}{4}\sin^22\theta_{12}c_{\theta}-(1-\frac{1}{4}\sin^22\theta_{12})c_{\theta}^2}{2-\frac{1}{4}\sin^22\theta_{12}c_{\theta}
+(1-\frac{1}{4}\sin^22\theta_{12})c_{\theta}^2},\label{fluxrat}
\end{eqnarray}
where we have used \eqref{iocosdel}, \eqref{iotheta23} and neglected terms of $\mathcal{O}(s^2_{13})$. It is evident from the approximate expressions \eqref{fluxrat} that in the exact $\mu\tau$ interchange limit $\theta=\frac{\pi}{2}$, all the flavor flux ratios converge to the value $\frac{1}{2}$. It is clear from the figure as well as from the approximate expression of $R_e$ that for $R_e<\frac{1}{2}(R_e>\frac{1}{2})$, we have $\theta<\frac{\pi}{2}(\theta>\frac{\pi}{2})$. Since \eqref{iotheta23} implies $2\theta_{23}=\pi-\theta$, observed value of $R_e$ will give a definite value of $\theta_{23}$. In particular, $\theta>\frac{\pi}{2}$ implies $\theta _{23}<\frac{\pi}{4}$ and vice versa. Similar conclusion can be made from the observed value of $R_\mu$. Although, the expression for $R_{\mu}$ in \eqref{fluxrat} is quadratic in $\cos {\theta}$, only one of the roots of this equation belongs to the numerically allowed range of $\theta$ (Table~\ref{osc2}). However, a definite observational value of $R_{\tau}$ cannot unambiguously predict the value of $\theta$. This is because of the quadratic dependence of $R_\tau$ on $c_\theta$ which is clearly visible from Fig.\ref{ReRmuRtau}, specifically for $\theta<\pi/2$. For consistency, the unique value of $\theta$ determined from the future precision measurement of $R_e$ (or $R_\mu$) lead to a theoretical prediction of the ranges of $R_\mu$ (or $R_e$) and $R_\tau$ which should in turn match the observed values of $R_\mu$ (or $R_e$) and $R_\tau$.
 Conversely, if $\theta_{23}$ is measured with significant precision in a complementary experiment (e.g. long baseline experiments), the range of each $R_l$ can be uniquely predicted for all $l$, which can again be compared with the observations in IceCube.

%
\section{Summary and conclusion} In this paper, we propose an invariance of the low energy neutrino Majorana mass term under a mixed $\mu\tau$-flavored CP symmetry $CP^{\mu\tau\theta}$ compounded with a generic Friedberg-Lee (FL) transformation on the left-handed flavor neutrino fields. Both types of mass ordering are allowed with a nondegenerate neutrino mass spectrum and vanishing value for the smallest neutrino mass as a direct consequence of FL invariance. While the atmospheric mixing angle $\theta_{23}$ is in general nonmaximal ($\theta_{23}\neq \pi/4$), the Dirac CP phase $\delta$ is exactly maximal ($\delta=\pi/2,3\pi/2$) for IO and nearly maximal for NO owing to $\cos\delta\propto \sin\theta_{13}$ though the deviation from maximality does not exceed $0.4^\circ$ on either side of the maximal value $\delta=3\pi/2$. It also turns out that one of the Majorana phases, $\alpha$, is restricted to lie at its CP conserving values while the other, $\beta$, admits a simple linear relation with $\delta$ leading to a tiny Majorana CP violation. For the IO, $\theta_{23}$ is, in general, nonmaximal but $\delta$ is maximal irrespective of the value of $\theta_{23}$. For the NO, the Majorana CP violation sneaking through the Majorana phase $\beta$ is numerically insignificant so that the model essentially predicts vanishing Majorana CP violation. Evidently, any large departure of $\delta$ from $3\pi/2$, will exclude our model. After fitting the neutrino oscillation global fit data, we also consider a numerical study of $\nu_\mu\rightarrow \nu_e$ oscillation which is expected to show up Dirac CP violation in different long baseline experiments. Finally, assuming purely astrophysical sources, we calculate the Ultra  High Energy (UHE) neutrino flavor flux ratios at neutrino telescopes such as IceCube. From this we comment on the predictability of the octant of $\theta_{23}$ in our model.

\acknowledgments

We thank Ambar Ghosal for bringing the Friedberg-Lee symmetry into our attention. R. Sinha is supported by the the Department of Atomic Energy (DAE), Government of India. SB is supported by Council of Scientific and Industrial Research (CSIR), Government of India. R. Samanta is supported by Newton International Fellowship from Royal Society (UK) and SERB (India).


\end{document}